\documentclass[twocolumn]{aastex7}

\usepackage[T1]{fontenc}
\usepackage{graphicx}	
\usepackage{float}
\usepackage{subcaption}
\usepackage{amsmath}	
\usepackage{amssymb}	
\usepackage{wrapfig}
\usepackage{natbib}
\usepackage{xcolor}
\usepackage{ulem}
\usepackage{soul}
\usepackage{hyperref}

\usepackage{nopageno}
\usepackage{geometry}
\usepackage{xcolor}
\definecolor{blazeorange}{rgb}{1.0, 0.4, 0.0}
\definecolor{seagreen}{rgb}{0.18, 0.55, 0.34}
\definecolor{rufous}{rgb}{0.66, 0.11, 0.03}
\definecolor{royalfuchsia}{rgb}{0.79, 0.17, 0.57}
\definecolor{scarlet}{rgb}{1.0, 0.13, 0.0}
\definecolor{royalpurple}{rgb}{0.47, 0.32, 0.66}
\definecolor{darkblue}{rgb}{0, 0, 0.66}

\newcommand{\dmunit}{pc/cm$^{3}$ }
\newcommand{\rmunit}{rad/m$^{2}$ }

\DeclareRobustCommand{\VAN}[3]{#2}
\let\VANthebibliography\thebibliography
\def\thebibliography{\DeclareRobustCommand{\VAN}[3]{##3}\VANthebibliography}

\hypersetup{linkcolor=blue,citecolor=blue,filecolor=cyan,urlcolor=magenta}
\usepackage{graphicx}   
\usepackage{amsmath}    
\usepackage{amssymb}    
\usepackage{newtxtext,newtxmath}
\usepackage{color}
\usepackage{threeparttable}
\usepackage{hyperref}
\usepackage{ulem}
\usepackage{mathtools}

\def\d{\mathrm{d}}

\newcommand{\lta}{\lower 2pt \hbox{$\, \buildrel {\scriptstyle <}\over {\scriptstyle \sim}\,$}}
\newcommand{\gta}{\lower 2pt \hbox{$\, \buildrel {\scriptstyle >}\over {\scriptstyle \sim}\,$}}

\definecolor{blazeorange}{rgb}{1.0, 0.4, 0.0}
\definecolor{seagreen}{rgb}{0.18, 0.55, 0.34}
\definecolor{rufous}{rgb}{0.66, 0.11, 0.03}
\definecolor{royalfuchsia}{rgb}{0.79, 0.17, 0.57}
\definecolor{scarlet}{rgb}{1.0, 0.13, 0.0}
\definecolor{royalpurple}{rgb}{0.47, 0.32, 0.66}

\begin{document}

\title{Constraining the near-source relativistic wind medium using Fast Radio Burst circular polarization data
}

\correspondingauthor{Om Gupta} 
\email{omgupta@utexas.edu}

\author[0000-0001-8470-7289]{Om Gupta}
\email{omgupta@utexas.edu}
\affiliation{Department of Astronomy, University of Texas at Austin, Austin, TX 78712, USA}

\author[0009-0003-8955-7402]{Pawan Kumar}
\email{pk@astro.as.utexas.edu}
\affiliation{Department of Astronomy, University of Texas at Austin, Austin, TX 78712, USA}

\author[0000-0001-7833-1043]{Paz Beniamini}
\email{pazb@openu.ac.il}
\affiliation{Department of Natural Sciences, The Open University of Israel, P.O Box 808, Ra'anana 4353701, Israel}
\affiliation{Astrophysics Research Center of the Open university (ARCO), The Open University of Israel, P.O Box 808, Ra'anana 4353701, Israel}
\affiliation{Department of Physics, The George Washington University, 725 21st Street NW, Washington, DC 20052, USA}

\begin{abstract}
Fast Radio Bursts (FRBs) exhibit diverse spectro-temporal characteristics, which can probe vital propagation and source physics via Stokes polarimetry. 
We investigate whether the circular polarization (Stokes $V$) observed in some bursts is produced by Faraday conversion in the near-source wind of magnetars rather than being intrinsic to the source. Our calculation includes the increase in the effective mass of $e^\pm$ in the presence of the FRB wave.

We find that Faraday conversion in the magnetar wind can explain the broad range of observed circular polarization in FRBs, including its frequent non-detection.
Observationally derived upper limits on $V$ provide stringent constraints on the wind luminosity, magnetization, bulk Lorentz factor, and effective particle mass when ions are present.
When available, frequency resolved Stokes spectra offer direct estimates of the wind environment. The Stokes parameters can undergo rapid oscillations with frequency in the high-wind/low-FRB-luminosity regime, resulting in Stokes-V depolarization. Bursts with significantly lower luminosities than typical FRBs can also develop measurable circular polarization, within the model framework. Additionally, separate zones are favored for significant circular polarization and rotation measure, when the model is applicable.

The model constrains instantaneous wind parameters for several sources, including FRB 20201124A, FRB 20180301A, and SGR 1935+2154. 
This work represents the first instance in which properties of winds from compact objects associated with FRBs are inferred from polarization data.
\end{abstract}


\section{Introduction}

Fast Radio Bursts (FRBs) are ultra-short duration radio transients that are ubiquitous across the sky. Thousands of them have been detected so far, with the population divided into repeaters and (apparent) non-repeaters. Telescopes with a large field-of-view, such as CHIME are excellent at detecting non-repeating bursts and finding repeaters \citep{CHIME2022}. On the other hand, sensitive telescopes such as FAST, and large bandwidth systems, such as the Parkes-UWL and the GBT-BL, provide excellent follow-ups of repeating sources \citep{Jiang2020, ParkesUWL2020, GBTBL2023}.

The polarization data collected by these instruments serves as a vital forensic probe of source emission properties, as well as the conditions of the medium along the propagation path. In pulsar astronomy, the S-shaped polarization angle swing is a well-known example that informs about the emission geometry (e.g., \citealt{Radhakrishnan1969}). The recent detection of a similar swing in an FRB \citep{Mckinven2025} suggests a polar cap origin for the emission \citep{BK2025a}.
A common example of signal modification beyond the source by the magneto-ionic medium the FRB traverses through, is the Faraday rotation of the linearly polarized component. Different levels of Faraday rotation are contributed by the near-source environment, the host galaxy, the inter-galactic medium, the Milky Way, and even the Earth's ionosphere. Under certain special conditions, linearly polarized radio waves can develop circular polarization via Faraday conversion, which is a higher order effect than Faraday Rotation. Faraday Conversion can be achieved when (1) the plasma shows birefringence, generally due to the presence of a background magnetic field, and (2) the magnetic field is tilted away from the direction of propagation. The strength of Faraday Conversion can be enhanced by changing the particle response to the electromagnetic wave using relativistic motion.

Circular polarization (abbreviated as CP hereafter) is a fairly common feature of many repeating FRBs, such as FRB 20201124A \citep{Hilmarsson2021, Jiang2022, Xu2022, Kumar2022}, FRB 20220912A \citep{Feng2024, Zhang2023, Ravi2023}, FRB 20240114A \citep{Tian2024}, FRB 20121102A \citep{Feng2022b}, FRB 20190520B \citep{Anna-Thomas2023}, and FRB 20180301A \citep{Price2019, Uttarkar2024}. Most bursts have small CP fraction $\Pi_V < 10\%$, but polarization statistics varies from source to source. For example, $\sim 27\%$ brights bursts in \citet{Xu2022} from FRB 20201124A exhibited $\Pi_V > 10\%$. The lack of circular polarization information for most of the non-repeating sample seems to be primarily due to the lack of detection pipelines which store full polarization information. Early recorded polarization results found that a good fraction of the bursts showed circular polarization \citep{Feng2022a}. Recent work with ASKAP \citep{Scott2025} and DSA \citep{Sherman2024} shows that a small fraction of non-repeating FRBs have negligible CP, while most bursts exhibit $\Pi_V < 0.2$. On the other hand, another non-repeating sample from Apertif/Westerbork yielded a small average CP fraction of $0.1\pm0.06$ \citep{Pastor-Marazuela2025}.

In this paper, we study the effect of Faraday conversion in the relativistic wind 
medium from the FRB central engine, most likely consisting of $e^\pm$ pairs, moving away from the source at relativistic speeds and extending to perhaps $10^{-2}\,\mathrm{pc}$ or more.
The presence of ions in the wind is also investigated. This changes the expected behavior of the polarization vector because in a purely $e^\pm$ plasma, Faraday rotation is zero.
With ions, the polarization vector exhibits a more complex evolution on the Poincare sphere.

The most notable part of our approach is the consideration of a relativistically intense FRB pulse propagating through a physically plausible magnetized wind medium. The treatment of the relativistically intense wave is performed through the relativistic mass increase of particles in the presence of such a wave. This treatment is naturally able to explain why only a few bursts have measurable circular polarization. In general, a higher magnitude of CP and furthermore, frequency-dependent CP oscillations require a higher wind luminosity, and lower wind magnetization, bulk Lorentz factor, and ion loading. With the ability to constrain and, in some cases, extract exact instantaneous wind properties, the model is implemented on available data for several FRBs. This work not only advances on previous theoretical approaches but also lays down a clear and verifiable set of predictions.

Here, we discuss relevant literature to put our results in context. Early work on Faraday conversion as applied to FRBs \citep{Vedantham2019} focused on constraining the properties of a two-zone model comprised of a synchrotron nebula and a cold-dense medium, such as the supernova remnant, confining it.
At that time, bursts from their test case, FRB 20121102A, had only been observed to be $\sim 100\%$ linearly polarized at high frequencies (e.g., \citealt{Gajjar2018}; see \citealt{Feng2022b, Feng2022a} for newer discoveries, and Section \ref{sec:plasma_scattering_discussion} for context).
Nevertheless, if the synchrotron nebula does not contribute to Faraday conversion, then the particle energy in the nebula must be extremely high \citep{Huang2011}.

\citet{Gruzinov2019} showed that Faraday conversion in cold non-relativistic plasma can become pronounced at magnetic field reversals.
The reversal points where maximum conversion happens have magnetic field lines perpendicular to the line of sight, which is a common feature in all Faraday conversion models, including ours. Their work predicts that $V \sim \omega_B {\rm RM}^{1/2} \lambda^2 / 2\pi c$, where $\omega_B$ is the angular cyclotron frequency (see Discussion \ref{sec:discussion_rm} for validity). 
Because the above expression for $V$ is proportional to $\omega_B/\omega \ll 1$, we expect bursts to exhibit a relatively small change in CP for a given change in RM. For all repeaters exhibiting CP, it is seen that $V$ changes significantly between bursts, irrespective of how RM is changing (e.g., \citealt{Hilmarsson2021, Xu2022}). This behaviour cannot be explained by Faraday Conversion at field reversals. While we argue the non-applicability of their model for a large number of bursts, we illustrate its success in context of a specific burst in Section \ref{sec:counter_examples}.

\citet{Lyutikov2022} considers Faraday conversion in a regime where the angular frequency of the wave in the wind frame $\omega' \ll \omega_B'$, which is the wind-frame cyclotron frequency. A brief discussion of the applicability of this regime is presented in Section \ref{sec:lyutikov2022}. A recent work by \citet{Wang2025} on both cold and relativistically hot magnetized plasma shows that circular polarization can arise in two ways. The first is by Faraday conversion, and the second is by free-free absorption of linear polarization. They calculate the properties of these hot and cold media, which exist as dense clouds. However, the location of these clouds is undetermined and they are proposed to move across the sightline, instead of being associated with non-transient structures such as the persistent radio source or the supernova remnant. 
With this freedom, \citet{Wang2025} are able to fit complex polarization spectra with ease. However, explaining high-magnitude flat $V$ spectra is challenging (see Section \ref{sec:data_analysis}). 

An independent way to obtain circular polarization has been discussed in \cite{BKN2022}. This work considers plasma scattering of FRB waves, and shows that under certain conditions, multi-path propagation can lead to depolarization and/or convert a fraction of the original wave polarization to a circular component. The effect is distinct from Faraday conversion discussed in the present work. It is stochastic in nature and has a distinct spectral dependence. It is also likely to be more prominent in FRBs which exhibit strong plasma scattering (see Section \ref{sec:plasma_scattering_discussion}).

The organization of this paper is as follows. First, we set up Faraday conversion in a relativistic magnetized pair wind in Section \ref{sec:basic_formalism}. In the same section, we extend the basic formalism to include realistic wind conditions in our numerical setup, following which Section \ref{sec:general_results} illustrates the general properties of our model. In Section \ref{sec:data_analysis}, we analyze FRB polarization data from several repeaters to understand their wind environments, as well as the limitation of our wind model in the context of one burst which cannot be explained with this model. Crucial aspects of this theoretical approach are discussed in Section \ref{sec:discussion}, with the main takeaways summarized in Section \ref{sec:summary}.

\section{Formalism for Faraday Conversion in a Relativistic pair plasma Wind} \label{sec:basic_formalism}

We start by considering a cold electron-positron wind, moving away from the FRB source with a Lorentz factor $\gamma$. $\gamma$ is considered to be constant for now, but will have a radial dependence in the numerical implementation. The equation for rotation of the polarization vector $\vec{P}$ on the Poincare sphere is 
\begin{equation}
    {d \vec{P}\over dr} = \vec{P}\times\vec{\Omega},
    \label{dP-dz1}
\end{equation}
where 
\begin{align}
    \vec{P} & \equiv (Q, U, V) \\
     \vec\Omega & =  -\eta_2 \sin^2\theta_B'\left[\cos2\phi_B \,{\bf\hat Q} + \sin2\phi_B\,{\bf\hat U}\right], \label{eq:Omega_def}  \\
     \eta_2 &\equiv {2\gamma^2\omega_p^2\omega_B^2\over \omega^3 c}, \quad \omega_p^2 \equiv {4\pi e^2 n\over m\gamma}, \quad \omega_B = {e B_{0\perp}\over m c\gamma}, \label{eta2-deflab} \\
     \hat{\bf B}_0 & = (\sin\theta_B \cos\phi_B, \sin\theta_B\sin\phi_B, \cos\theta_B).
\end{align}
In the above equations, $n$ is the total number of electrons and positrons per unit volume, ${\vec{B}}_0$ is the unperturbed magnetic field in the medium, $\hat{\bf k}_0$ is the wave propagation direction, $\cos\theta_B \equiv \hat{\bf B}_0 \cdot \hat{\bf k}_0$, and $\eta_2$ is the Faraday conversion coefficient, with all quantities in the observer/lab frame. See Appendix \ref{app:poltransfer_transform} for the transformation of comoving-frame quantities (denoted as primed, e.g., $\theta_B'$) to the lab frame. Because we are considering an electron-positron plasma, the \( \hat{\bf V} \) component of \( \vec{\Omega} \) is exactly zero regardless of the orientation of \( \vec{B}_0 \). Note that all non-zero subscripts denote normalized quantities, i.e., $X_{10} = X/10^{10}$.

We have taken the Stokes parameter $I$, which represents the wave intensity, to be unity. Therefore, $(Q,U,V)$ above are the usual Stokes parameters divided by $I$. It should be noted that the normalized  Stokes parameter \( V \) and the combination \( Q^2 + U^2 \) are Lorentz invariant (see, e.g., \citealt{Cocke1972}). This is an unsurprising result, as all observers should agree on the degree of linear and circular polarization of an electromagnetic pulse. However, the polarization angle of the linearly polarized component is a frame-dependent quantity.

Without loss of generality, we can take $\hat{\bf k}_0 = \hat{\bf z}$ (where $\hat{\bf z}$ is along the radial direction $r$), and the background magnetic field is toroidal, with $\theta_B=\pi/2$, yielding  $\hat{\bf B}_0 \cdot \hat{\bf k}_0=0$. We choose $\hat{\bf B}_0 =\hat{\bf x}$, i.e. $\phi_B = 0$. With these choices, $\vec\Omega = -\eta_2 \hat{\bf Q}$ and the equation for evolution of Stokes parameters simplifies considerably and is given by
\begin{equation}
    {dU\over dr} = -\eta_2 V, \quad {dV\over dr} = \eta_2 U.
    \label{dP-dz2}
\end{equation}
These equations demonstrate that Faraday rotation is identically zero in an $e^\pm$ plasma, but Faraday conversion (which is otherwise typically smaller) is not. They can be combined to yield
\begin{equation}
    {d^2 U \over d\zeta^2} = - U, \quad {\rm where}\quad \zeta(r)\equiv \int^r_{R_0} d\tilde{r}\, \eta_2(\tilde{r}).
    \label{dP-dz3}
\end{equation}
Equation \ref{dP-dz3} has the solution 
\begin{equation}
    U = A\cos(\zeta+a),\quad V = A\sin(\zeta+a).
    \label{sol1-UV}
\end{equation}
The constants $A$ and $a$ are determined from the initial condition for the wave as it enters the wind medium at $R_0$. We consider $R_0$ to be the light cylinder distance, which marks the end of the magnetospheric region and the beginning of the wind region. $R_0$ can be related to the magnetar spin period as $R_0 = cP_{\rm spin}/2\pi = (10^{11} \;{\rm cm}) (P_{\rm spin}/20.94\;{\rm s})$. For instance, for a 100\% linearly polarized wave at $R_0$ (where $\zeta=0$), $a=0$ and $A$ is determined by the position angle for the wave at $R_0$; if the electric vector is oriented at $\pi/4$ with respect to the x-axis at $R_0$, then $Q(R_0)=0$, $U(R_0)=1$, and $V(R_0)=0$. We can see that $A=1$. For the same magnetic field orientation and 100\% linearly polarized wave, when the electric field is not necessarily oriented at $\pi/4$ relative to $\hat{\bf x}$, $A=U(r=R_0)\equiv U_0$, and that sets the maximum circular polarization that can be produced by Faraday conversion. If $V_0 \neq 0$, then $A = \sqrt{U_0^2 + V_0^2}$. The general case of arbitrary magnetic field orientation is discussed in Section \ref{sec:num_model}.

The parameter $\eta_2 \propto r^{-4}$, since the plasma density and magnetic field strength decrease with distance from the central engine as $r^{-2}$ and $r^{-1}$, respectively, when the circumburst medium consists of a steady wind from the FRB source (see e.g., \citealt{Kirk2009}).
Using the definition of the wind luminosity,
\begin{equation}
    L_w = 4\pi r^2 v \gamma m c^2 n (1+\sigma_w),
    \label{eq:wind_luminosity_defn}
\end{equation}
where, $v \sim c$, and the definition of the wind magnetization parameter,
\begin{equation}
    \sigma_w \equiv {B_0^2\over 4\pi n m c^2\gamma},
\end{equation}
the plasma and cyclotron frequencies in the lab frame can be written as follows,
\begin{equation}
    \omega_B = {e B_0\over mc\gamma} = (3.2\times 10^7 {\rm rad\, s}^{-1})\, {L_{w,37}^{1/2}\over r_{13} \gamma} \sqrt{\sigma_w\over 1+\sigma_w}.
    \label{omegaB-B0}
\end{equation}
and 
\begin{equation}
    \omega^2_p \equiv {4\pi e^2 n\over m\gamma} = {\omega^2_B \over \sigma_w}.
        \label{omegap}
\end{equation}
Therefore,
\begin{equation}
    \eta_2 \equiv {2\gamma^2\omega_p^2\omega_B^2\over \omega^3 c} = (7\times 10^{-11} {\rm cm}^{-1}) {L_{w,37}^2\sigma_w \over r_{13}^4 (1+\sigma_w)^2\omega_{10}^3\gamma^2}.
    \label{eta2a}
\end{equation}
For non-dissipative acceleration of a magnetized wind, the quantity $(1 + \sigma_w)\gamma$ remains constant with radius; that is, as the wind Lorentz factor increases, $\sigma_w$ decreases to keep their product fixed, thereby conserving the jet power. 

Another important consideration for Faraday conversion of high amplitude EM pulses, such as FRBs, is that e$^\pm$ exposed to the wave are set into motion at relativistic speeds. This modifies the equations we have thus far presented. Let us define the dimensionless amplitude of EM pulse to be $\alpha$ as follows
\begin{equation}
    \alpha \equiv {e E_w\over m c\omega} = 3.2\, L_{\rm frb,43}^{1/2} r_{13}^{-1}\omega_{10}^{-1},
    \label{eq:alpha}
\end{equation}
where $E_w=\sqrt{\frac{L_{\rm frb}}{cr^2}}$ \& $\omega$ are wave electric field and frequency respectively, and $L_{\rm frb}$ is the radio luminosity of an FRB. Light charged particles interacting with this wave acquire a transverse Lorentz factor $\gamma_\alpha \sim (1 + \alpha^2)^{1/2}$. In the lab frame, the total Lorentz factor $\gamma_{\rm tot} \equiv \gamma \gamma_\alpha$, so the particles' effective mass in the lab frame becomes $m \gamma_{\rm tot}$. 
As a result, the cyclotron and plasma frequencies are suppressed by an additional factor of $\gamma_\alpha$ and $\gamma_\alpha^{1/2}$, respectively as compared with Eqns (\ref{omegaB-B0}) and (\ref{omegap}).

The propagation of a nonlinear electromagnetic (EM) pulse through a magneto-ionized medium, and the corresponding evolution of the Stokes parameters with distance, is inherently complex. We propose an ansatz\footnote{This assumption may require verification of the validity of polarization transfer through detailed analytical and numerical calculations, even though the relativistic mass increase affecting $\omega_p$ and $\omega_B$ is a well-known laser plasma physics phenomenon (see e.g., \citealt{Cattani2000, Sharma2012, Saedjalil2016, Sano2017}).} that $\eta_2$ is reduced by a factor of $\gamma_\alpha^3$, since $\eta_2 \propto \omega_p^2 \omega_B^2$ for EM waves with $\alpha \ll 1$ (see Eq. \ref{eta2-deflab}). Under this prescription, the modified expression for $\eta_2$ becomes:
\begin{eqnarray}
    \eta_2 \sim 
    (7\times 10^{-11} {\rm cm}^{-1}) {L_{w,37}^2 \sigma_w \over (1+\sigma_w)^2\gamma^2}\left\{ \begin{array}{ll}
      {1\over 35\, r_{13} L_{frb,43}^{3/2}}   & \quad \quad  \alpha \gta 1 \\ \\
      {1 \over r_{13}^4 \omega_{10}^3}  & \quad \quad \alpha \lta 1 
    \end{array}\right.
    \label{eq:eta2_cases}
\end{eqnarray}

\begin{figure}[h]
\centering
\includegraphics[width=0.47\textwidth]{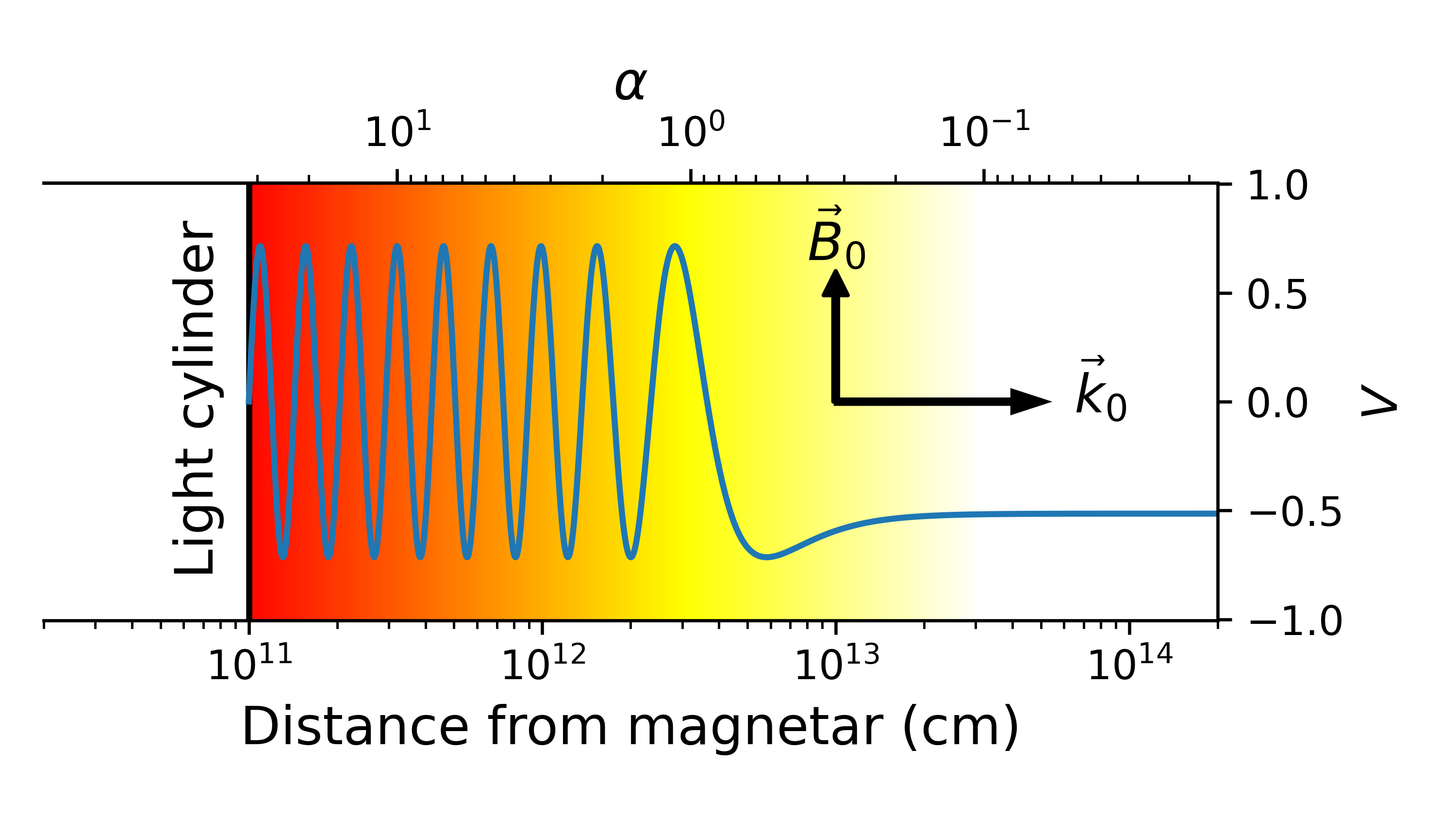}
\caption{The circular polarization $V$ as a function of distance and the EM strength parameter $\alpha$ (shown on the top, and also as the color gradient) for an FRB with $L_{\rm frb}=10^{41}{\rm erg\, s}^{-1}$. Propagation within the magnetosphere is not considered in this work. The background magnetic field $\vec{B}_0$ is taken to be perpendicular to $\vec{k}_0$ outside the light cylinder. The effect of the nonlinear FRB wave amplitude is incorporated by reducing the Larmor frequency by a factor $\alpha$. The initial polarization is: $Q(R_0)=0.7$, $U(R_0) = [1-Q(R_0)^2]^{1/2}$, and $V(R_0)=0$. The parameters for this calculation are: $\omega = 10^{10}$ rad s$^{-1}$, $L_w =10^{37}$ erg s$^{-1}$, $R_0 = 10^{11}$ cm, $\sigma_w=10$, and $\gamma=10$.
}
\label{fig:StokesV_schematic}
\end{figure}

\begin{figure*}
    \centering
    \includegraphics[width=0.48\linewidth]{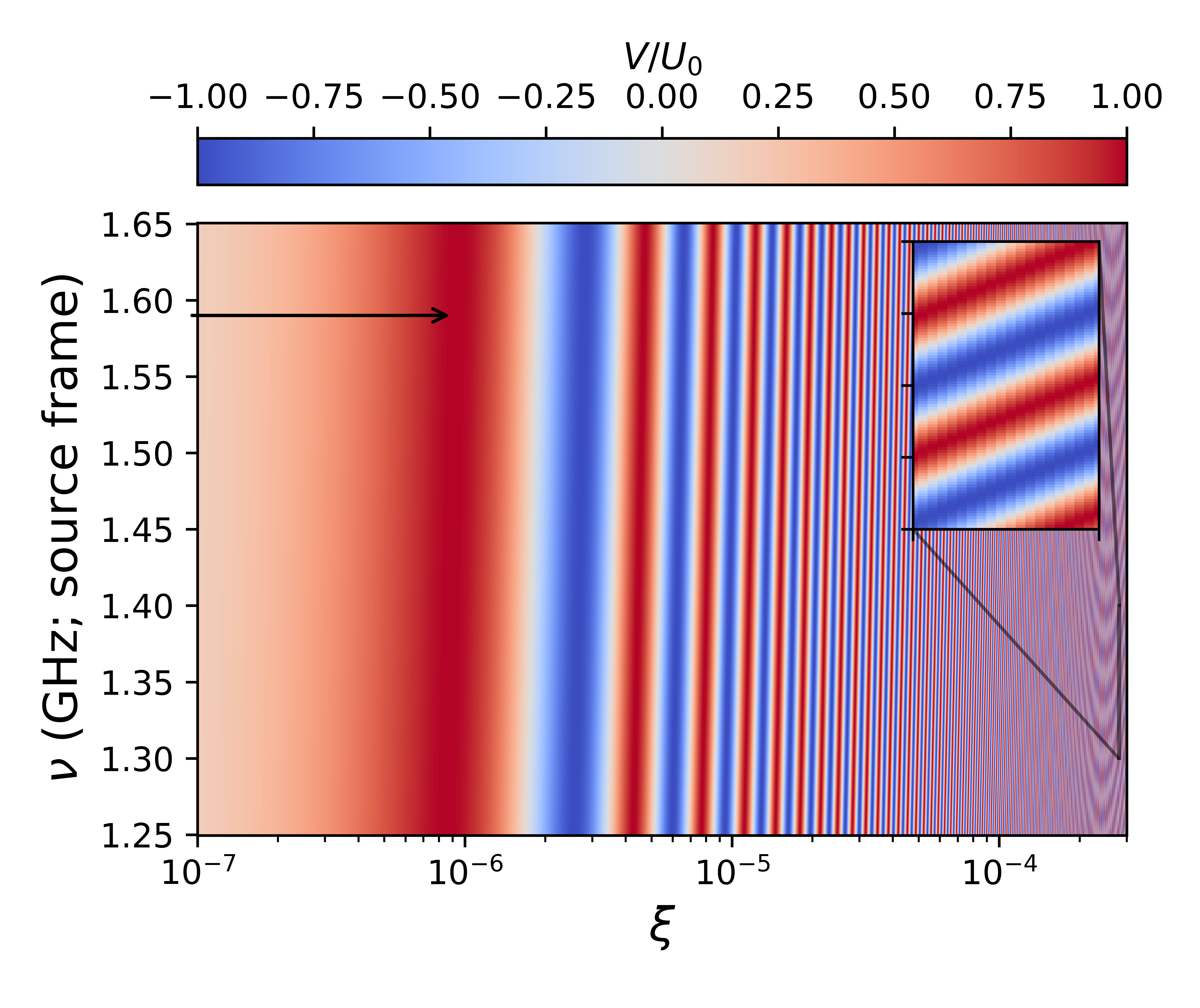}
    \includegraphics[width=0.48\linewidth]{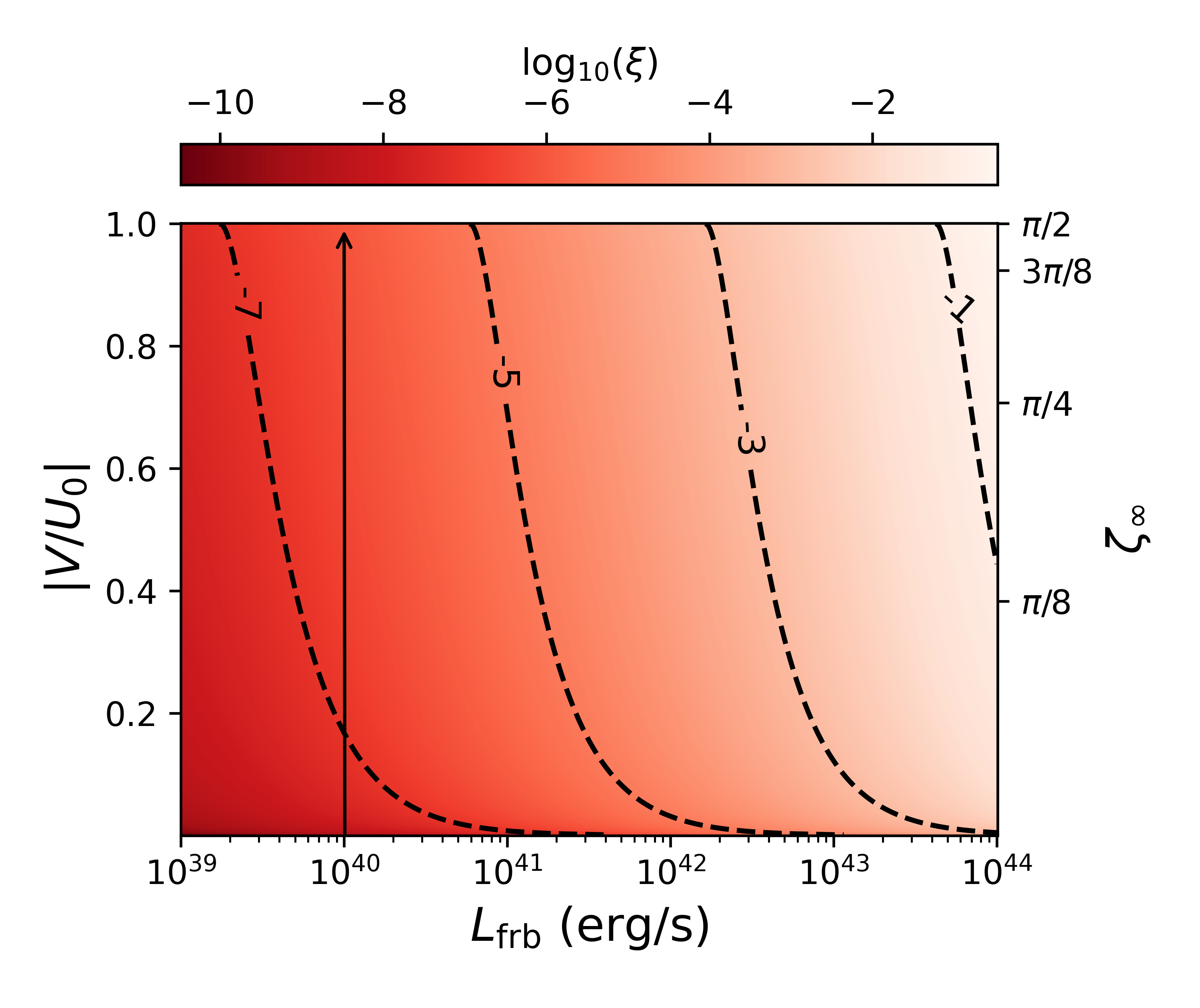}
    \caption{(\textit{left}) The frequency dependence of $V/U_0$ (in color) from our analytical model as a function of the wind parameter $\xi \equiv L_{w, 37}^2\sigma_w / (1+\sigma_w)^2 \gamma^2$. The frequency-dependent oscillation of $V$ intensifies with increasing $\xi$, and the inset shows rapid oscillations from a small patch at high $\xi$. $L_{\rm frb} = 10^{40}$ erg/s, and the FRB is initially 100\% linearly polarized, with $U_0 = 1.0, Q_0 = V_0 = 0$ at the base of the wind $R_0$. Only $U_0$ is converted into circular polarization, since—by the choice of coordinates—the magnetic field in the $e^\pm$ wind is aligned along the x-axis. Inset coordinates are $\xi \in [2.8-2.82]\times10^{-4}$ and $\nu \in [1.3-1.4]$ GHz. (\textit{right}) The color denotes $\xi$ as a function of the FRB luminosity and the observed $|V|$ normalized by $|U_0|$, computed when $\zeta_{\infty} \leq \pi/2$ (shown on the right-side y-axis). The dashed contour lines indicate levels of log$_{10}(\xi)$ corresponding to the color. The angular frequency $\omega = 10^{10}$ rad/s, and input Stokes parameters are the same as the left plot. The two plots can be related as follows: the vertical arrow in the right plot at $L_{\rm frb} = 10^{40}$ erg/s is equivalent to the horizontal arrow in the left plot at $\nu \approx 1.59$ GHz, extending to the first region where $V/U_0=1$.}
    \label{fig:Lfrb_nu_vs_xi}
\end{figure*}

Since $\eta_2$ scales as $r^{-4}$ at large radii where $\alpha(r) < 1$, the integral of $\eta_2$ over radius converges rapidly and thus, the smallest radii in this regime dominate the contribution to Faraday conversion. At small radii where $\alpha > 1$, the integrand scales as $1/r$, and the asymptotic value of $\zeta$, denoted as $\zeta_\infty$, exhibits a logarithmic dependence on $R_0$. This asymptotic value, provided that $\alpha(R_0) > 1$, is
\begin{equation}
    \zeta_\infty \equiv \int_{R_0}^\infty dr\, \eta_2(r)\approx {20\,\xi\over  L_{\rm frb,43}^{3/2}}\bigg[\log(r_{\alpha}/R_0)+0.35\bigg],
    \label{chi-inf1}
\end{equation}
where
\begin{equation}
    r_{\alpha}=(3.2\times 10^{13} {\rm cm}) {L_{\rm frb,43}^{1/2}\over \omega_{10} },
\end{equation}
denotes the radius at which the dimensionless nonlinearity parameter $\alpha$ for the FRB radio pulse decreases to approximately unity, and 
\begin{equation}
    \xi \equiv {L_{w,37}^2\sigma_w\over (1+\sigma_w)^2\gamma^2}.
    \label{eq:xi_defn}
\end{equation}
The evolution of CP as a function of distance in the wind is highlighted in Figure \ref{fig:StokesV_schematic}, which also illustrates how $\alpha$ varies in the region of interest. We only consider FRB propagation in the wind, and any magnetospheric contribution is outside the scope of this paper. This issue is discussed further in Section \ref{sec:magnetospheric_propagation}. The dependence of CP on $\xi$ and frequency, via Eqs (\ref{sol1-UV}) and (\ref{chi-inf1}), is shown in the left panel of Figure \ref{fig:Lfrb_nu_vs_xi}.

So far, we have treated the case of $r_\alpha > R_0$. What happens when $r_\alpha < R_0$? This situation can arise for low luminosity bursts, with $L_{\rm frb} < 10^{38} R_{0,11} \omega_{10}$ erg/s. The equation for $\zeta_\infty$ when $r_\alpha < R_0$ is obtained by using equation~(\ref{eta2a}) for $\eta_2$:
\begin{equation}
    \zeta_\infty \sim 2.33\times 10^8\; {\xi \over R_{0,11}^3 \omega_{10}^3}.
    \label{eq:chiinf_nonrel}
\end{equation}
Here, only the $\eta_2 \propto r^{-4}$ contribution is applicable in the wind region, and $\zeta_\infty$ drops fast with the distance of the starting point. In spite of this, Eq (\ref{eq:chiinf_nonrel}) shows that a measurable Faraday conversion can be obtained in the wind. 
When $\zeta_\infty \ll 1$, it is seen that $V \propto \lambda^3$. We shall see in Section \ref{sec:data_analysis} that $\xi \ll 10^{-8}$ for SGR 1935-2154, representing the typical values expected for Galactic radio sources.

\subsection{Constraints on wind associated with the FRB source from non-detection of circular polarization}
The degree of circular polarization due to Faraday conversion, as the FRB pulse exits the magnetar wind, is determined from equation~\ref{sol1-UV}, i.e. $V(r = \infty) = U_0 \sin(\zeta_\infty)$. For $\zeta_\infty < 1$, the degree of circular polarization is approximately $U_0 \zeta_\infty$, and it depends linearly on the magnetar wind properties through the parameter $\xi$. In this case, the absence of circular polarization from a burst immediately leads to a constraint on $\xi$.

For a given FRB with an observed luminosity $L_{\rm frb}$ and an upper limit on its CP $\lesssim V$, suggesting\footnote{$|V| \ll 1$ does not necessarily imply that $\zeta_\infty \ll 1$. Indeed, $\zeta_\infty = n\pi + \epsilon$ is consistent with a small value of $V$, where $n$ is an integer, and $\epsilon \ll 1$. However, this is more fine-tuned, requiring spectrum matching of $V$, which is considered later in this work.} that $\zeta_\infty \ll 1$, we can place a limit on the wind properties as follows:
\begin{equation}
    {L_{w,37}\, \sigma_w^{1/2}\over (1+\sigma_w)\gamma} \lesssim {[|V|/U_0]^{1/2} L_{\rm frb,43}^{3/4}\over 5}, 
    \label{Lw-limit}
\end{equation}
provided that $r_\alpha > R_0$. In Eq (\ref{Lw-limit}), we have ignored the $\log(r_\alpha/R_0)$ term present in Eq (\ref{chi-inf1}) in square brackets, to obtain a simple estimate. It can decrease the constraint on $L_w$ by up to a factor of 2.

As an example, for an FRB generated within the magnetosphere with $L_{\rm frb} = 10^{40}$ erg s$^{-1}$ whose observed degree of circular polarization is small, $|V/U_0| \lesssim 0.1$, while the linear polarization fraction is close to 100$\%$, we find, using Eq (\ref{Lw-limit}), that the luminosity of the magnetar wind is well constrained to be $L_{w} \lesssim 4 \times 10^{35}\;{\rm erg/s} \; (\sigma_w/100)^{1/2} (\gamma/10)$. The constraint on the wind parameters becomes increasingly severe as the FRB luminosity and $|V|$ decrease. 
These conditions are also realized if the medium just ahead of the FRB pulse, within hundreds of AU of the central object, is not the quiescent wind from the magnetar but plasma ejected during the explosion that triggered the FRB.

\subsection{Consideration of realistic wind conditions} \label{sec:num_model}
We examine next the ways in which more realistic wind conditions can be accommodated in our model. We include a mixture of protons (used interchangeably with ions) in the wind, as well as radial evolution of wind properties. These changes are implemented in our numerical model, and their effects are examined in Section \ref{sec:impact_realistic_conditions}.

\subsubsection{Ions in the wind}

The presence of ions in the wind from the central compact object has an important effect on converting a linearly polarized signal to being circularly polarized. $\Omega_V$ is no longer necessarily zero, as taken in Eq (\ref{eq:Omega_def}), and that can reduce the conversion of linear to circular polarization. The $\vec{V}$-component of $\vec\Omega$ is
\begin{equation}
    \Omega_V = {2\psi_p\gamma \omega_p^2\omega_B \cos \theta_B'\over c\omega^2 }\equiv \eta_1 \cos \theta_B'
    \label{eq:eta1_OmegaV}
\end{equation}
where $\psi_p$ is the number of protons per electron in the wind, $\cos \theta_B' \equiv B'_{0r}/B'_0$ is the radial component of the wind magnetic field in the wind frame ($\hat{\bf r}$ is also the direction of wave propagation), and the Faraday rotation coefficient defined fundamentally in the comoving frame $\eta_1' = \omega_p'^2\omega_B'^2 / c\omega'^2$. The transformation from its fundamental comoving-frame definition is given in Appendix \ref{app:poltransfer_transform}, while the explicit form of $\eta_1$ in the $\alpha \gtrsim 1$ and $\alpha \lesssim 1$ regimes is given in Appendix \ref{app:eta1}. 

The plasma frequency is also modified for a mixed $e^\pm$ and proton plasma and is given by:
\begin{equation}
    \omega_p^2 = {m\over m_E} \times {\omega_B^2\over\sigma_w}, \quad {\rm where}\quad {m_E\over m}\equiv  \frac{(2 -\psi_p)m + \psi_p m_p}{(2 -\psi_p)m}.
    \label{eq:omegap_effective_mass}
\end{equation}
$m_p$ is the proton mass, $m_E$ is the effective wind particle mass, and we have assumed that protons have the same bulk Lorentz factor as electrons and positrons. Appendix \ref{app:effective_mass} shows how this relation arises. The treatment of this mixed species plasma can be simplified by assuming a perpendicular magnetic field in the wind, which leads to $\eta_{2,\rm mixed} = \eta_2m/m_E$ and $\Omega_V=0$. The decrease by the factor $m_E/m$ is extended to Eqs (\ref{chi-inf1}), (\ref{eq:chiinf_nonrel}) and (\ref{eq:eta1_OmegaV}) as well. For an electron-proton plasma, $m_E \approx m_p$. Consequently, one requires a higher $L_{w,37} \sigma_w^{1/2} / (1+\sigma_w)\gamma$ to offset the smaller $m/m_E$ in order to achieve the same result as $\xi$ with $m_E = m$. We define an effective $\xi$ combining these parameters in Section \ref{sec:impact_realistic_conditions}.

\subsubsection{Radially changing wind properties}

In the ideal MHD regime, highly magnetized outflows accelerate quickly. Their Poynting to particle energy flux ratio $L_w/\dot{M}c^2 = \gamma(1+\sigma_w)$, which can also be shown to be the wind's conserved energy flux, thus remains constant in the absence of dissipation \citep{Michel1969, Okamoto1978}. Initial Lorentz factors of $\gamma_0$ and magnetization $\sigma_{w0}$, will attain final values of $\gamma_{\rm fms} = (\gamma_0 \sigma_{w0})^{1/3}$ and $\sigma_{w, \rm fms} = (\gamma_0 \sigma_{w0})^{2/3}$ respectively, by the time particles cross the fast-magnetosonic point or the Dicke-Alfven point\footnote{This is a critical point where the flow velocity reaches the Alfven velocity $= B_r/\sqrt{4 \pi \rho \gamma}$. Acceleration happens until this point because the magnetic pressure acting as the driving force is dynamically coupled to the central engine.}, $R_A = \sigma_{\rm w,fms} R_0$ \citep{Michel1991_book, Lyutikov2022}, where we take $R_0 = 10^{11}$ cm as the launch point of the wind. Acceleration is considered to be linear with radius, i.e., $\gamma(r) \propto r$ \citep{Contopoulos2002}. 

We next discuss the determination of $\cos \theta_B'$ and its dependence on $r$. The field $B_0(r)$ is given by equation (\ref{omegaB-B0}). The radial component of the field in the wind comoving frame ($B_{0r}' = B_0'\cos \theta_B'$) is the same as the lab frame value. However, the transverse component of the field in the wind frame is smaller than in the lab frame by a factor $\gamma_0$. Furthermore, the radial dependence of the two components is: $B_{0r} \propto 1/r^2$ and $B_{0\perp} \propto 1/r$ \citep{Kirk2009}. If we assume that $B_{0r}/B_{0\perp}=\mu$ at the launch radius ($R_0$) of the wind, then at larger radii $B'_{0r}/B'_{0\perp} = \mu \gamma_0 (R_0/r)$. Using this relation, we can calculate 
\begin{equation}
    \cos \theta_B' = {B'_{0r}(r) \over B_0'(r)} \sim {\mu\gamma_0(R_0/r)\over  \left[ 1 + \mu^2\gamma_0^2 (R_0/r)^2 \right]^{1/2}}.
    \label{eq:hat_B0r_prime}
\end{equation}
We take $\mu\gamma_0\sim 1$ for the numerical calculations, but $\mu$ can be affected via field reorganization caused by magnetar activity. For an aligned rotator, $\mu \gg 1$ near the magnetic axis, but for a misaligned rotator, its value would be lower.

The evolution of Stokes parameters with distance is determined using the following coupled equations:
\begin{equation}
 \begin{aligned}
    &{dQ\over dr}  = \eta_1 \cos \theta_B' U + \eta_2 \sin^2 \theta_B' \sin 2\phi_B\, V, \\
    &{dU\over dr} = -\eta_1 \cos \theta_B' Q -\eta_2 \sin^2\theta'_B \cos 2\phi_B\, V, \\ 
    &{dV\over dr} = \eta_2 \sin^2\theta'_B \cos 2\phi_B\, U - \eta_2 \sin^2 \theta_B' \sin 2\phi_B\, Q.
    \label{dP-dz4}
 \end{aligned}
\end{equation}
where $\sin\theta_B'\equiv |{\bf \hat{B}'_0\times \hat{k}_0'}|$. For larger $\mu \gamma_0$, $\cos \theta_B'(R_0)$ becomes closer to 1. Furthermore, it leads to a faster decline of $\cos \theta_B'$ with radius, but it happens farther in the wind. In Eq (\ref{dP-dz4}), higher $\cos \theta_B'$ means slower conversion to and from $V$, and more conversion between $Q$ and $U$ if $\psi_p > 0$.

\section{General results of the Faraday Conversion model} \label{sec:general_results}

In this section, we summarize the general behavior of circular polarization predicted by our model. We first describe the analytical results, highlighting how the single parameter $\xi$ controls both the magnitude and spectral structure of V. We then discuss how finite frequency resolution can lead to depolarization when $\xi$ is large. Finally, we present the modifications introduced by the numerical wind model, including radial evolution and ion loading, which lead to the definition of an effective wind parameter $\xi_E$.

\subsection{Wind parameter $\xi$ as the primary diagnostic}

In the analytical model, the level of circular polarization and its spectrum for a given FRB is determined by $\xi$. In other words, this parameter controls the cumulative strength of Faraday conversion in the wind. It is degenerate between combinations of $L_w$, $\sigma_w$ and $\gamma_0$. 
From Eqs (\ref{eq:wind_luminosity_defn}) and (\ref{eq:xi_defn}), small $\xi$ corresponds to a wind that has a small particle inertia and/or wind magnetization, and vice-versa.
The spectrum of circular polarization as a function of $\xi$ is shown in Figure \ref{fig:Lfrb_nu_vs_xi}, for a particular FRB luminosity. Negligible CP is expected at low values of $\xi$; as $\xi$ increases, circular polarization develops, but more importantly, the value of $\xi$ dictates the strength of spectral oscillations of $V$ (also see the inset of Figure \ref{fig:Lfrb_nu_vs_xi}). Consequently, we interpret $\xi$ as the state of the wind, which when relaxed, results in no circular polarization, and which when extreme, results in rapid spectral variation of $V$. The bandwidth over which $|\Delta (V/U_0)| = 1$ is obtained by differentiating Eq (\ref{sol1-UV}) and substituting Eq (\ref{chi-inf1}), to get,
\begin{equation}
    \Delta \nu \approx 0.5\mbox{ MHz} \frac{L_{\rm frb, 41}^{3/2} \nu_9}{\xi_{-1} \cos \zeta_{\infty}}.
\end{equation}
The underlying assumptions in this approximation are that $\Delta \nu \ll \nu$, and consequently, also that $\cos \zeta_\infty \neq 0$. Most telescopes which can record FRB Stokes-$V$ polarization data can resolve this bandwidth, unless $\xi \gg 0.1$.

\subsection{Finite frequency resolution and spectral depolarization of CP} \label{sec:fractional_polarization}
\begin{figure*}
    \centering
    \includegraphics[width=0.48\linewidth]{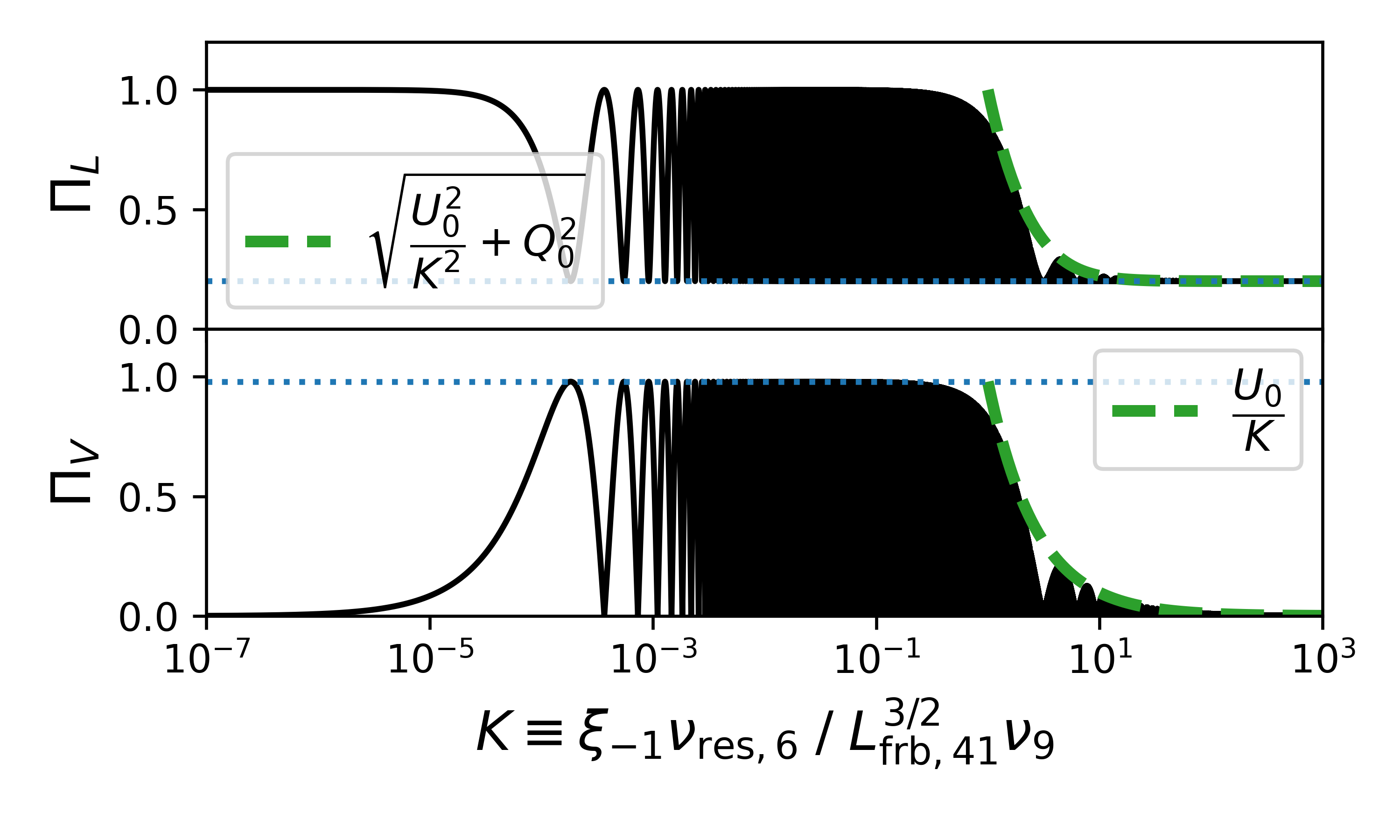}
    \includegraphics[width=0.48\linewidth]{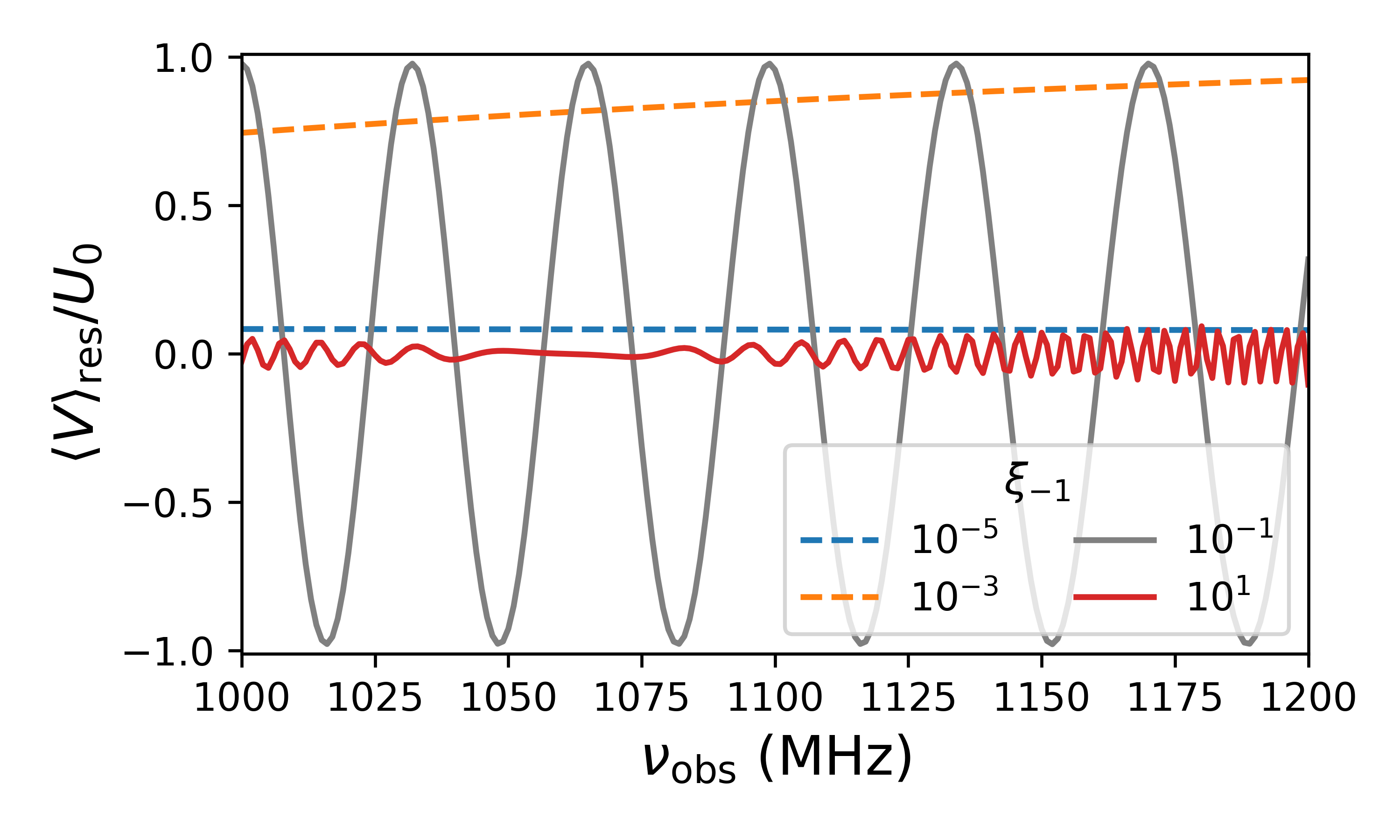}
    \caption{(\textit{left}) The evolution of the degree of linear polarization, $\Pi_L$, and the degree of circular polarization, $\Pi_V$, as a function of $K = \xi_{-1} \nu_{\rm res,6} \;/\; L_{\rm frb,41}^{3/2} \nu_9$, for the analytical case with $Q_0= 0.2$ and $V_0=0$. Because of the magnetic field orientation we have chosen, Faraday conversion between Stokes $U$ and $V$ occurs. For $K < 1$, the oscillations qualitatively follow a horizontal trajectory in Figure \ref{fig:Lfrb_nu_vs_xi}. For larger $K$, the envelope of LP $= \sqrt{(U_0/K)^2+Q_0^2}$ while that of CP $= U_0/K$ (both shown as the dashed green line), denoting the observed linear and circular depolarization. The dotted blue line in the upper panel shows $\Pi_L (K \gg 1) \rightarrow Q_0$, while in the lower panel, it denotes $U_0$.
    (\textit{right}) The evolution of $V/U_0$, averaged over a frequency resolution of $\nu_{\rm res} = 1$ MHz, denoted as $\langle V \rangle_{\rm res}/U_0$, as a function of frequency. We set $L_{\rm frb} = 10^{41}$ erg/s. The dashed and solid colored lines show a few cases of interest, and $\xi_{-1}$ values much higher or lower than those shown result in $\langle V \rangle_{\rm res}/U_0 \approx 0$ ($\Pi_V \approx 0$; which are not shown). The frequency evolution of these cases can be used to determine how strong $K$ (or $\xi$) is, except when $K$ is too small or when complete depolarization happens ($K$ is large).
    }
    \label{fig:piV_vs_K_nu_analytical}
\end{figure*}

One of the most common scenarios encountered in the literature is the lack of frequency-dependent polarization information, and the availability of Stokes parameters averaged over both time and frequency. These numbers summarize the polarization properties of each burst in a sample, and often represent a lack of human resource to process a large sample of bursts. Several oscillations can occur within the observing band of the detector such that the average circular component can appear depolarized. This case is discussed later in Section \ref{sec:impact_realistic_conditions} using the numerical approach. In the extreme case, several oscillations can happen within individual frequency channels of the detector, and which is much simpler to treat in the analytical case. 
The fractional CP, defined as $\Pi_V \equiv | \langle V \rangle_{\rm res}|$ is derived in Appendix \ref{app:fractional_polarization}. Here, the subscript $\langle \rangle_{\rm res}$ represents an average over the frequency resolution of the detector. Under specific cases, the envelope of the oscillating function $\Pi_V$ in Eq (\ref{eq:avgV_res}) can be simplified to,
\begin{align}
    \Pi_V^{\rm env} \approx U_0 \times
    \begin{cases}
        \sin \zeta_\infty &{\rm for}\; \dfrac{\xi_{-1} \nu_{\rm res,6}}{L_{\rm frb, 41}^{3/2}\nu_9} \lesssim 10^{-3}\\
        1 &{\rm for}\; 10^{-3} \lesssim \dfrac{\xi_{-1} \nu_{\rm res,6}}{L_{\rm frb, 41}^{3/2}\nu_9} \lesssim 1 \\
        \dfrac{L_{\rm frb, 41}^{3/2}\nu_9}{\xi_{-1} \nu_{\rm res,6}} \quad &{\rm for}\; \dfrac{\xi_{-1} \nu_{\rm res,6}}{L_{\rm frb, 41}^{3/2}\nu_9} \gtrsim 1 .
    \end{cases}
    \label{eq:CP_envelope}
\end{align}
These cases can be clearly distinguished in Figure \ref{fig:piV_vs_K_nu_analytical}, and simply outline that the CP fraction is initally negligible and then increases.
$\Pi_V^{\rm env}$ then becomes comparable to $U_0$, but actual CP fraction can oscillate within this envelope depending on the exact value of $\xi$. As the frequency-dependent oscillations become too rapid, $\Pi_V^{\rm env}$ starts declining as $1/\xi$. The right panel of Figure \ref{fig:piV_vs_K_nu_analytical} shows that observationally, the wind conditions can still be inferred in several cases by checking how $\Pi_V$ varies with frequency.

\subsection{The impact of realistic wind conditions} \label{sec:impact_realistic_conditions}
The numerical scheme introduces a non-zero proton fraction, radial evolution of wind properties, and a radially evolving magnetic field structure, as described in Section \ref{sec:num_model}.
Here, we perform analytical examinations on the radial evolution of different properties to assess their impact when implemented in the numerical model. 

The evolution of $\vec{\Omega}$, about which $\vec{P}$ rotates on the Poincare sphere also informs the predictions of this model. A schematic of the Poincare sphere is shown in the left panel of Figure \ref{fig:delta_radius}. The angle $\phi_B$ is assumed to be constant in the wind, which means that the azimuth of $\vec{\Omega}$ is fixed. The movement of $\vec{\Omega}$ is restricted to be along a longitude of the Poincare sphere, and its latitude is determined by the value of $\eta_1$. Appendix \ref{app:omega_evolution_poincare} shows that the extent of the movement of $\vec{\Omega}$ is governed by $\delta \propto \psi_p \sqrt{L_{\rm frb, 43} / L_{w,37}}$ when $\sigma_w \gg 1$ and $\alpha \gg 1$. For typical parameters, $\vec{\Omega}$ gradually settles in a direction almost aligned with the $Q-U$ plane, as the magnetic field becomes orthogonal to the propagation direction. This has an important consequence after passing through the wind, that $\langle V \rangle \rightarrow 0$ when computed over several spectral oscillations. 

Only for a heavily ion-loaded wind, close to being an electron-proton plasma, can the asymptotic value of $\delta$ be large enough that the spectral oscillations of $V$ are asymmetric about 0. According to Eq (\ref{eq:delta_cases}), the evolution of $\delta$ is frequency independent for $\alpha\gg 1$. Furthermore, $\delta \propto \omega_{10}$ for $\alpha \lesssim 1$, such that changes in the frequency evolution are evident only over several$\times 100$ MHz, and may not be readily distinguishable over typical detector bandwidths. The bottom line is that $\langle \vec{P}\rangle_{\rm res} \rightarrow \vec{\Omega}(r \rightarrow\infty)$, where the relation increasingly holds as the frequency resolution of the detection pipeline becomes poorer.

So far, we have presented the results of radially constant wind properties. The radial evolution of $\sigma_w$ and $\gamma$ is discussed in Appendix \ref{app:understand_radial_evolution}, with the simplified case of a transverse magnetic field. It is shown that the evolving case shows slower frequency-dependent $V$ oscillations than the non-evolving case, for the same $\xi$, with $\Delta \omega_{\rm evol} / \Delta \omega_{\rm non-evol} = \gamma_{\rm fms} \gamma_0^{-1}$.

\begin{figure*}
    \centering
    \includegraphics[width=0.48\textwidth]{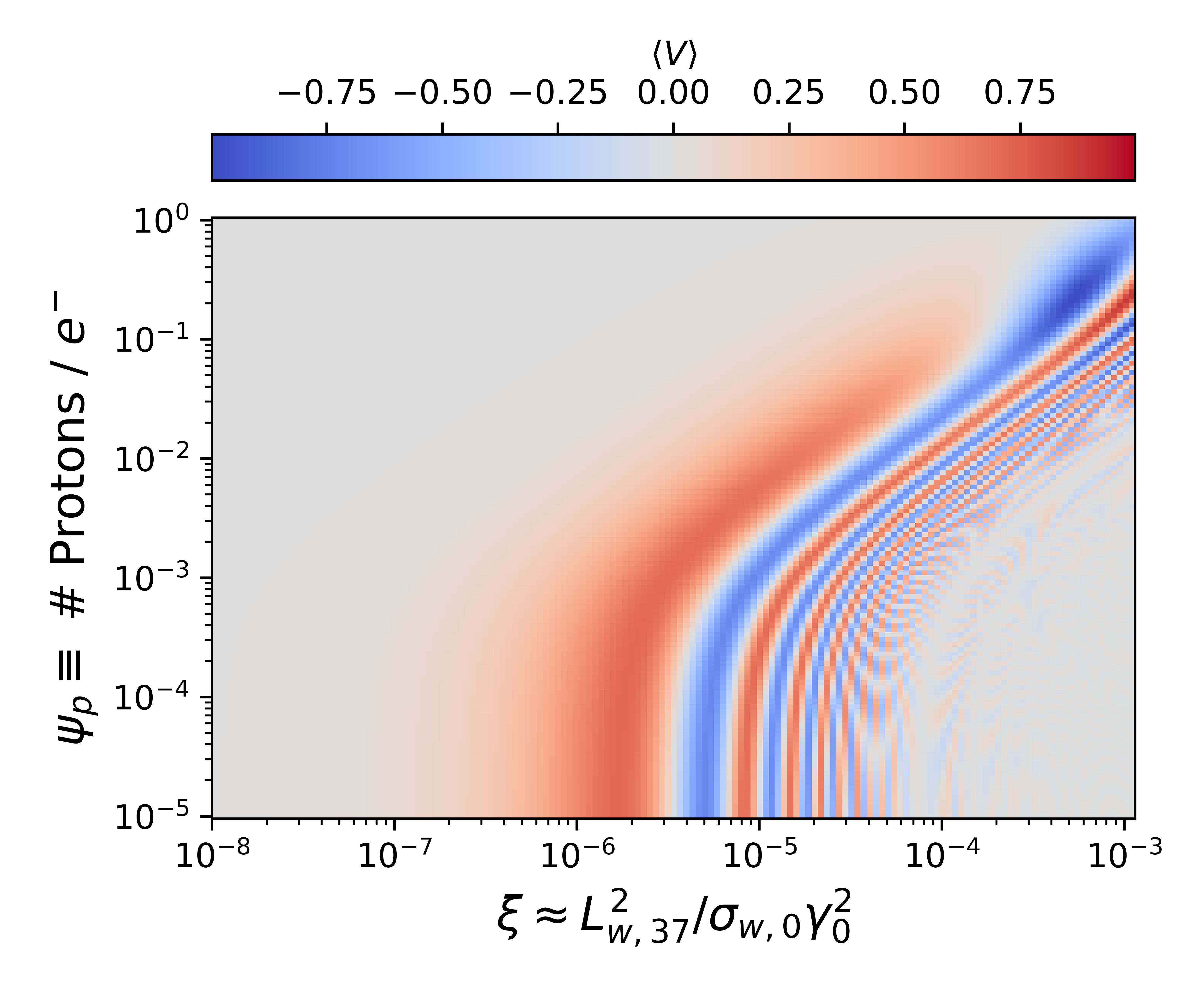}
    \includegraphics[width=0.48\textwidth]{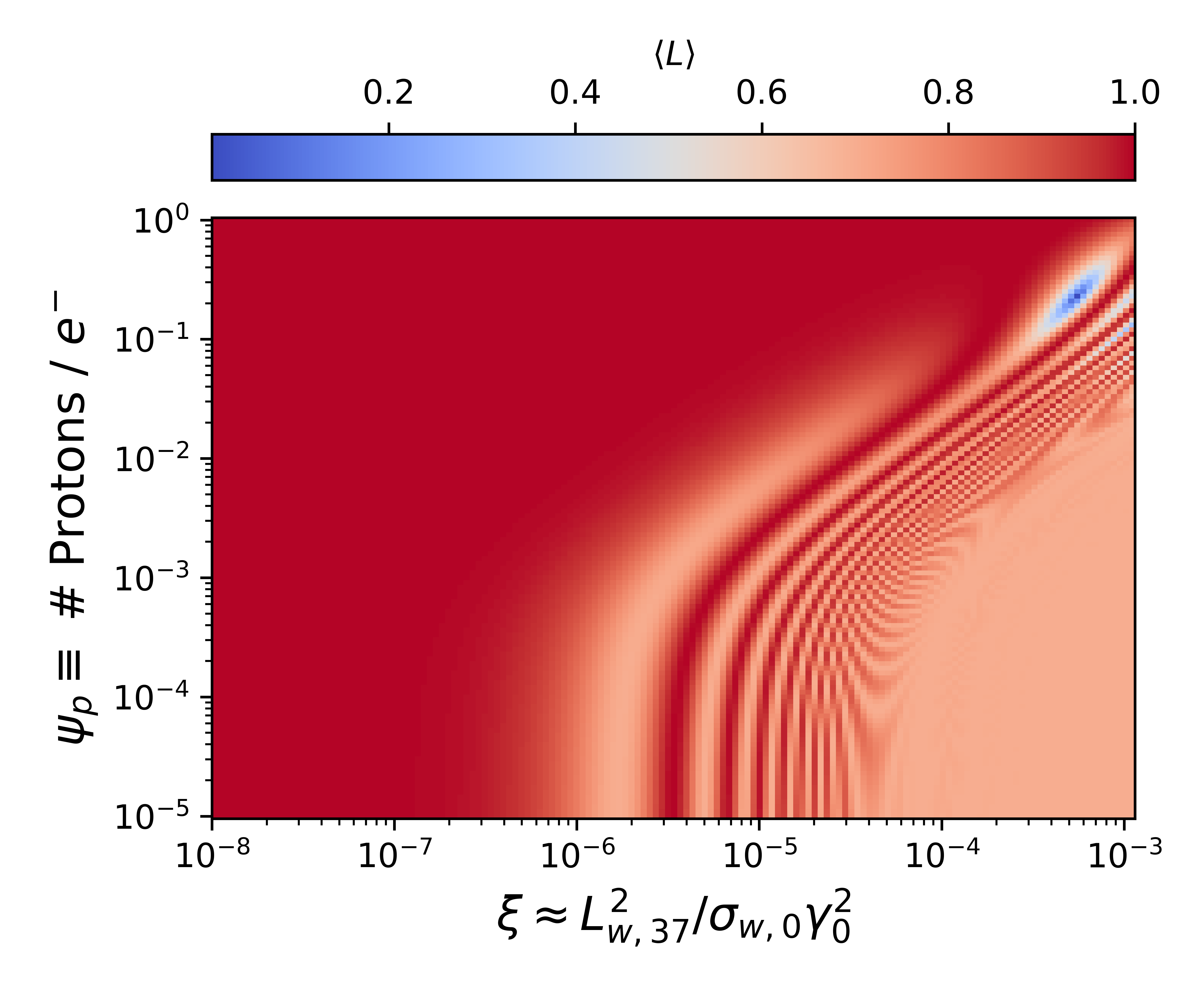}
    \caption{The averaged (\textit{left}) circular $\langle V \rangle$, and (\textit{right}) linear polarization $\langle L \rangle$ (in color) over the frequency range of $1.3-1.6$ GHz, as a function of the ion loading $\psi_p$ and the wind parameter $\xi$. 
    The negligible change in $\xi$ when moving downwards in the plot where $\psi_p<10^{-3}$, indicates that $\langle V \rangle$ asymptotically approaches the electron-positron pair plasma value. The diagonal trend at higher $\psi_p$ emphasizes the interplay between $\xi$ and $m/m_E$ to maintain a constant $\xi_E$. Rapid  spectral variation causes $V$-depolarization seen at higher $\xi$. While $V$ depolarizes, the oscillations of $L$ average to a fixed value $= [\Omega_Q^2(r \rightarrow \infty) + \Omega_U^2(r \rightarrow \infty)]^{1/2}$. The full numerical model is implemented here including the radial evolution of $\gamma$ and $\vec{B}_0$. $L_{\rm frb} = 10^{40}$ erg/s, and input Stokes parameters $Q_0 = 0.7$, $U_0 = 0.714$, and $V_0 = 0$.
    }
    \label{fig:Lfrb_iload_vs_xi}
\end{figure*}

As discussed in Section \ref{sec:num_model}, the presence of ions in the wind leads to the definition of an effective $\xi$,
\begin{equation}
    \xi_E \equiv {m\over m_E} \xi \approx \frac{L_{w,37}^2 m}{\sigma_{w0} \gamma_0^2 m_E}.
    \label{eq:xi_numerical}
\end{equation}
Due to radial evolution of the wind environment in the numerical model, however, the spectral dependence of $V$ is no longer uniquely described by a single value of $\xi_E$. If the constituent wind properties are specified, alternate combinations with slightly different $\xi_E$ can also fit the same spectral variations. This complication arises solely because of the radial dependence of wind variables. In Figure \ref{fig:Lfrb_iload_vs_xi}, we show the average circular polarization over 300 MHz bandwidth as a function of the ion loading and $\xi \approx L_{w,37}^2/\sigma_{w0} \gamma_0^2$, which are the two components of $\xi_E$. It also shows that with rapid spectral oscillations at large $\xi$, $\langle V \rangle$ becomes close to zero. 

In summary, $\xi$ is the primary indicator of the strength of Faraday Conversion in the analytical model, while $\xi_E$ is the key indicator in a more realistic wind environment. Furthermore, finite frequency resolution can suppress observable CP fraction when propagation-induced frequency-dependent oscillations are strong.

\section{FRB Circular Polarization Data and Its Implications for Circum-Burst Medium Properties} \label{sec:data_analysis}

We focus next on the important features of the analysis done on existing FRB data in Appendix \ref{app:analysis}, and present them in a concise manner. For the inferred wind properties based on each burst, see Appendix \ref{app:analysis}.

In our model, frequency evolution and its amplitude together, not just the magnitude of $V$, make the correct marker of wind conditions. The analysis of burst 521 from FRB 20201124A (see middle panel of Figure \ref{fig:frb20201124a_Jiang2024_Kumar2022}) challenges the notion that frequency-independent (or very slow frequency dependence) large $\Pi_V$ is unsupported by Faraday conversion \citep{Jiang2024}. Burst 521 shows $\langle V \rangle = -0.909$, but has close to no frequency evolution across the observed bandwidth, and the inferred $\xi=-5.02$ (Table \ref{tab:frb20201124a_gam0V0}). 
This low $\xi$ corresponds to weak cumulative conversion and therefore slow spectral evolution (e.g., see left panel of Figure \ref{fig:Lfrb_nu_vs_xi}). Burst 1472 of \citet{Xu2022}, with the highest average $\Pi_V$ ($\langle V \rangle = -0.751$) in their FRB 20201124A sample, shows a similar behavior and may be easily fit by our model.

Previous theories using Faraday Conversion \citep{Gruzinov2019, Wang2025} cannot explain the high-magnitude almost flat $V$-spectrum. 
It can be partially explained when the linear polarization component is absorbed significantly more than CP in an arbitrarily high density magnetized transient cloud \citep{Wang2025}. Even then, initial CP is needed, and the resultant polarization shows small-amplitude spectral oscillations around the mean, which are possibly detectable.
Our model does not require intrinsic CP to explain this behavior, whereas previous works require both fine-tuning and intrinsic occurrence.

Burst-to-burst variation in $\xi$ may imply temporal changes in wind properties. Using spectral information for FRB 20201124A, we derive accurate ranges for plausible $\xi$ values in the left panel of Figure \ref{fig:frb20201124a_Jiang2024_Kumar2022}. Although, this is limited to a small sample of mostly non-consecutive bursts, it provides us with an approximate state of the wind during the observation window, of $\xi \lesssim 4.9\times 10^{-4}$.
Even with limited data in the form of pulse-averaged CP values, we show for FRB 20220912A how its wind environment evolves over a timescale of $\sim 5000$ s (see Figure \ref{fig:frb20220912a_xi_slice_3panel}). Many bursts are used for this analysis, assuming some simplifying conditions. We see that $\xi \sim 10^{-5} - 10^{-7}$, but some bursts require $\xi$ up to $6\times 10^{-4}$.

\begin{figure*}
    \centering    \includegraphics[width=0.48\linewidth]{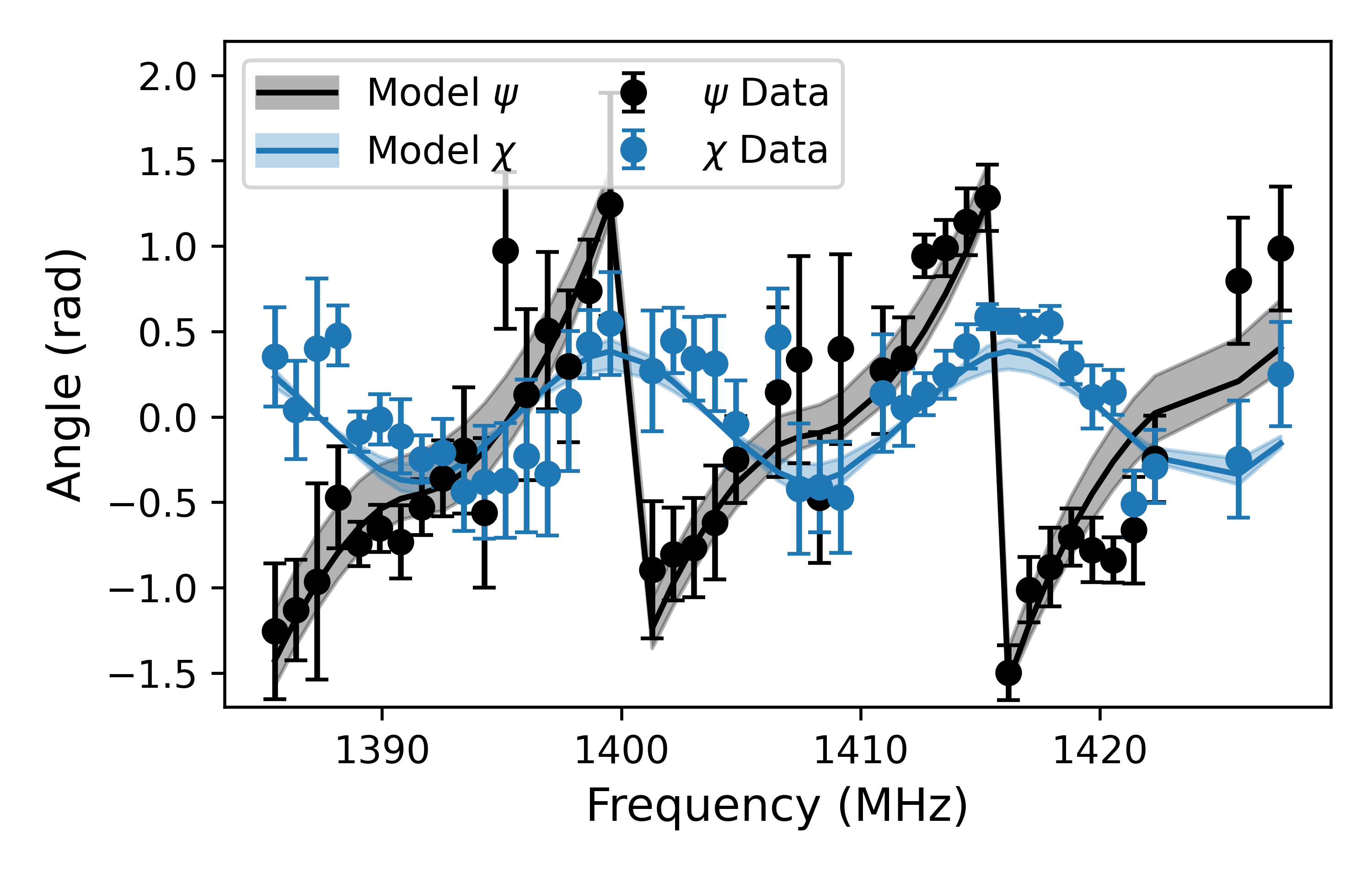}
    \includegraphics[width=0.48\linewidth]{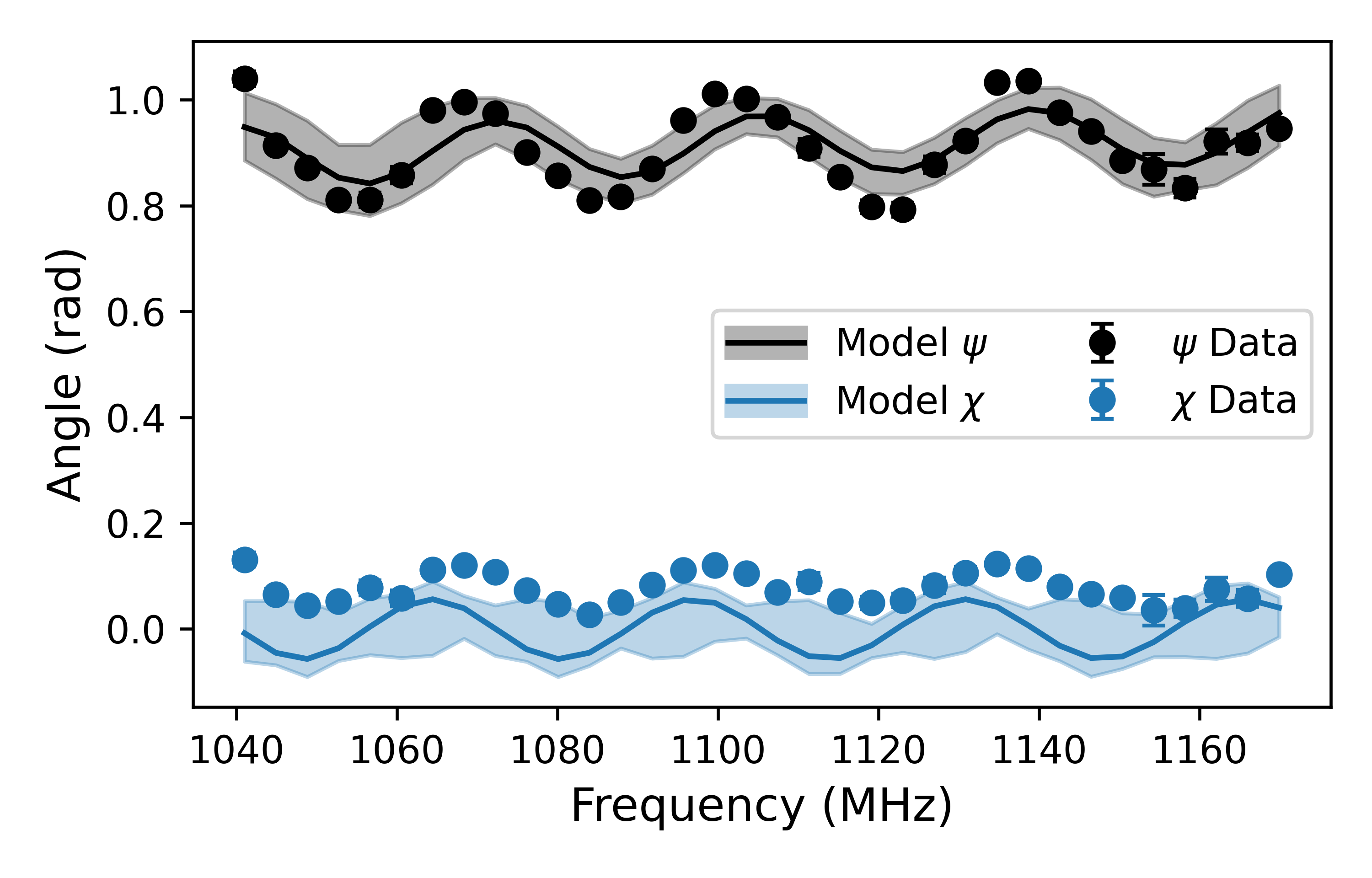}
    \caption{(\textit{left}) Model fit of the frequency dependence of the observed Faraday uncorrected polarization angle, $\psi = 0.5 \tan^{-1}(U/Q)$ (in black), and ellipticity angle, $\chi = 0.5 \tan^{-1}(V/\sqrt{Q^2+U^2})$ (in blue), of the discovery burst of FRB 20180301A \citep{Price2019}. We choose $L_{w,37} = 100$ and restrict the initial CP by setting $\chi_0 = 0$. The best-fit RM is found to be $-3235$ \rmunit. The shaded $68\%$ confidence region shown is obtained using the methodology described in Appendix \ref{app:frb20201124a}. Our Faraday conversion model can reproduce closely the most rapid frequency-dependent variations seen for any FRB to date.
    (\textit{right}) Model fit of the frequency evolution of $\psi$ (Faraday corrected) and $\chi$ of FRB 20201124A's burst 926 from \citet{Xu2022}, allowing full freedom on $\psi_0$ and $\chi_0$, and choosing $L_{w,37}=8$.
    This model is not able to explain the observed circular polarization, because it is limited to producing $\chi(\nu)$ that is symmetric about $0$ for this burst's luminosity, $L_{\rm frb} \sim 6.9 \times 10^{40}$ erg/s.
    }
    \label{fig:uttarkar2024_xu2022}
\end{figure*}

From the analysis of the FRB 20180301A discovery burst polarization data showing rapid frequency evolution \citep{Price2019, Uttarkar2024}, we find that our model can explain the observed properties alongside an RM close to what has been originally estimated (observed RM$_{\rm obs} = -3163$ \rmunit, our model's best-fit RM$_{\rm fit} = -3235$ \rmunit). This suggests re-examining the applicability of the Generalized Faraday Rotation (GFR) model \citep{Uttarkar2024, lower2021}, which obtains a very different value of RM$_{\rm GFR} = 28$ \rmunit. The fit using our model is shown in Figure \ref{fig:uttarkar2024_xu2022} (left panel). \citet{Wang2025} also point out that GFR should not be able to extract the underlying physical conditions, but accept and use the GFR-derived RM to infer the properties of the intervening medium. Their approach cannot inherently determine if the proposed RM is correct or not.

\begin{figure}
    \centering
    \includegraphics[width=\linewidth]{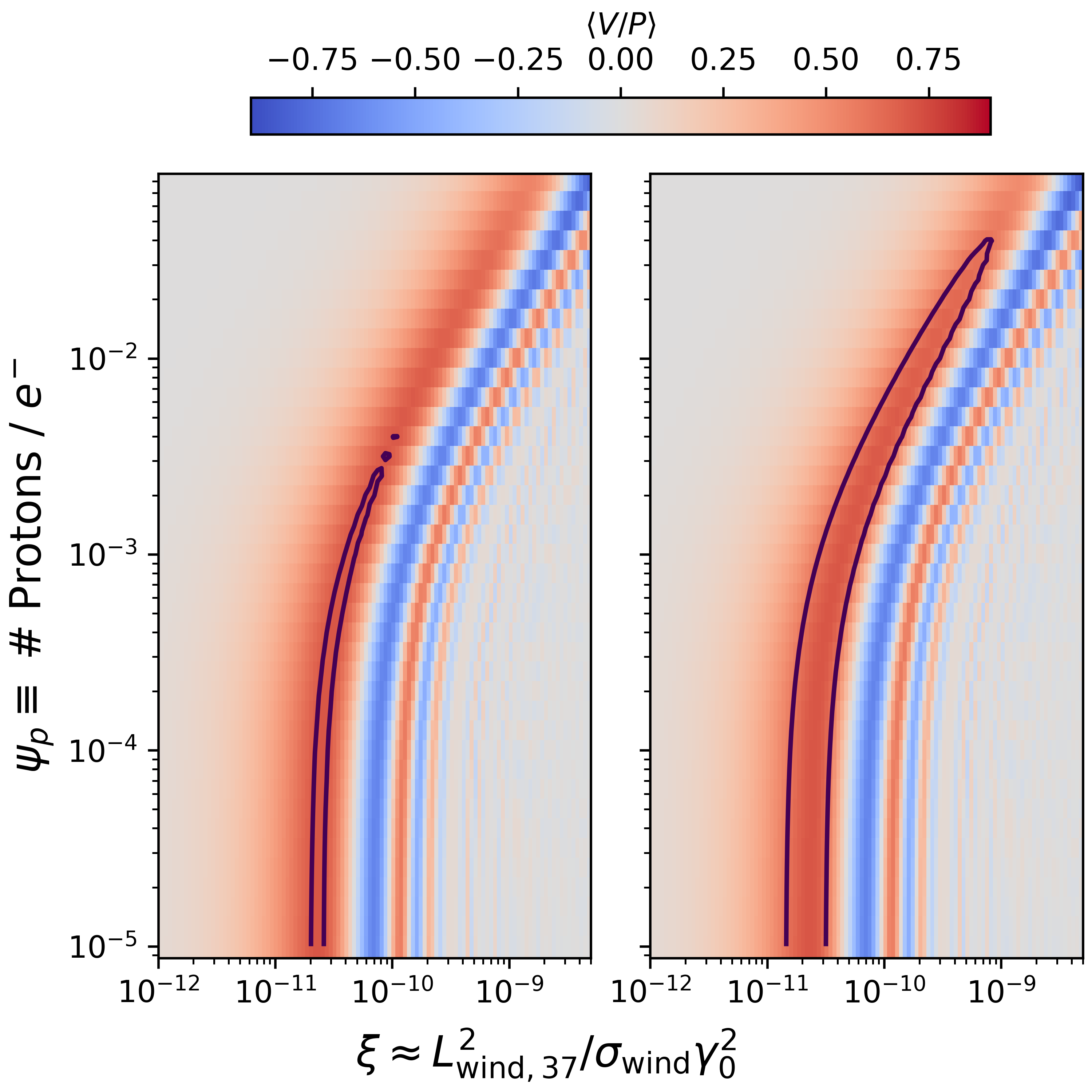}
    \caption{Contour lines (in black) show the parameter space enclosing the averaged circular polarization $\langle V/P \rangle\pm0.1$ over the observed frequency range $1260 - 1388$ MHz, for SGR 1935+2154 radio bursts from \citet{Kirsten2021}. Here, $P$ refers to the total polarization of the bursts, since it is not $100\%$. The background plot is similar to Figure \ref{fig:Lfrb_iload_vs_xi}, but for the burst luminosities corresponding to the (\textit{left}) first and (\textit{right}) second bursts that were separated by about 1.4s. The numerical model is initialized with $R_0 = 1.55 \times 10^{10}$ cm, which is the light cylinder radius of SGR 1935+2154.}
    \label{fig:sgr1935_contour}
\end{figure}

We also apply our model in the $r_\alpha < R_0$ case with radio bursts from SGR 1935+2154 (\citealt{Kirsten2021}; alternatively FRB 20200428). These bursts have orders of magnitude lower luminosity than typical FRBs, but we have shown that significant CP can be obtained even in this case (see Eq \ref{eq:chiinf_nonrel}). The contour of the burst-averaged circular polarization showing the possible range of parameter space is shown in Figure \ref{fig:sgr1935_contour}. This is the only observation where a rough constraint on the proton loading can be identified, showing $\psi_p \lesssim 0.01$. It is clear from the plots that any non-negligible level of CP over the observed bandwidth can only be sustained in a narrow range of $\xi \sim 10^{-11} - 2\times 10^{-10}$. Both higher and lower $\xi$ yield negligible $V$, the first due to depolarization and the second due to intrinsically weak wind conditions. The reader is advised caution in interpretation since spectral information is not used.

\subsection{A Nonconforming Burst and Model Limitations: Burst 926 of FRB 20201124A from \citet{Xu2022}} \label{sec:counter_examples}

Among all bursts recorded from FRB 20201124A, this burst exhibits the most rapidly varying spectrum. The model outlined in this work is unable to fit the spectrum, and the data with the best-fits are shown in Figure \ref{fig:uttarkar2024_xu2022} (right panel), for the case of $L_{w, 37}=8$.

The inability to fit this burst arises from one of the specifications of the wind model, that the wind magnetic field becomes increasingly toroidal with radial distance. Consequently, the vector $\vec{\Omega}$, dictating the polarization transfer in Eq. (\ref{dP-dz1}), becomes almost co-planar with the $Q-U$ plane. Further evolution of the polarization state involves revolutions around $\hat{\bf \Omega} = (\cos 2\phi_B, \sin 2\phi_B, \delta(r\rightarrow \infty))$, with $\delta(r \rightarrow \infty) \ll 1$ typically. This results in the oscillation of V being symmetric about zero. Oscillations of $V$ which are not symmetric about 0 can be generated for an electron-proton plasma ($\psi_p \sim 1$) and $L_{\rm frb} \lesssim 10^{39}$ erg/s, but not for this burst.

Apart from intrinsic polarization, a plausible explanation could be an ion-rich significant magnetar outburst which reorganizes the magnetic fields up to large distances in the wind. Magnetospheric Faraday conversion may also be possible, but it must explain the observed $V/I$ statistics and simultaneously address complex propagation dynamics (see Section \ref{sec:magnetospheric_propagation}).

\citet{Xu2022} empirically show that the frequency-dependent oscillation follows a $\lambda^2$ behavior. Such a wavelength dependence has been predicted for field reversals, and the Fourier transform of the $V$ frequency-spectrum can yield information about each reversal \citep{Gruzinov2019}. In \citet{Xu2022}, this burst's Stokes-$V$ is fit well with a single sinusoidal curve, which indicates that a dominant field reversal is the primary contributor. 
It can be shown easily that an initial non-zero circular polarization with a flat frequency spectrum of Stokes-$V$, can explain the vertical offset. This is valid with the assumption that the initial linear polarization is large.

\section{Discussion} \label{sec:discussion}

\subsection{Faraday Conversion in FRBs generated via the external shock model} \label{sec:external_shock}

The discussion thus far has implicitly considered that FRB radio waves are produced inside the light-cylinder radius. However, if the waves are generated far outside via synchrotron maser operating behind a blast wave shock front, then the emergent waves are primarily X-modes (e.g., \citealt{Beloborodov2020}). In this case, the electric field is perpendicular to the ambient magnetic field, and it can be easily seen that the Stokes parameters for the pulse at the source are: $Q_0=-1$, $U_0=0=V_0$. Hence, $\vec{P}$ at the source is parallel to $\vec{\Omega}$. Therefore, from equation (\ref{dP-dz1}), we find  that $\vec{P}$ does not change as the wave propagates through the magnetar wind, i.e. there is no conversion of linear polarization to circular polarization provided that the magnetic field orientation is independent of $r$ upstream of the shock front throughout most of the wind medium, the wavevector being perpendicular to the ambient magnetic field, and the wind composed entirely of e$^\pm$. 

This result is obtained from the polarization transfer of emission from a single coherent patch. However, the external shock is a large surface propagating outwards with time. Emission from varying transverse distances as the shock propagates has the same travel time to the observer. As such, superposition of multiple finite bandwidth wavefronts happens at each instant along the temporal profile of the FRB. In an $e^{\pm}$ wind, the above result of zero CP development remains valid. However, when $\psi_p > 0$, conversion of the initial X-mode to CP for each path along with the incoherent superposition of waves at the observer happens. A detailed study of this process is left to future endeavors.

\subsection{Constraints on the range of plausible wind parameters}
\begin{figure}
    \centering
    \includegraphics[width=\linewidth]{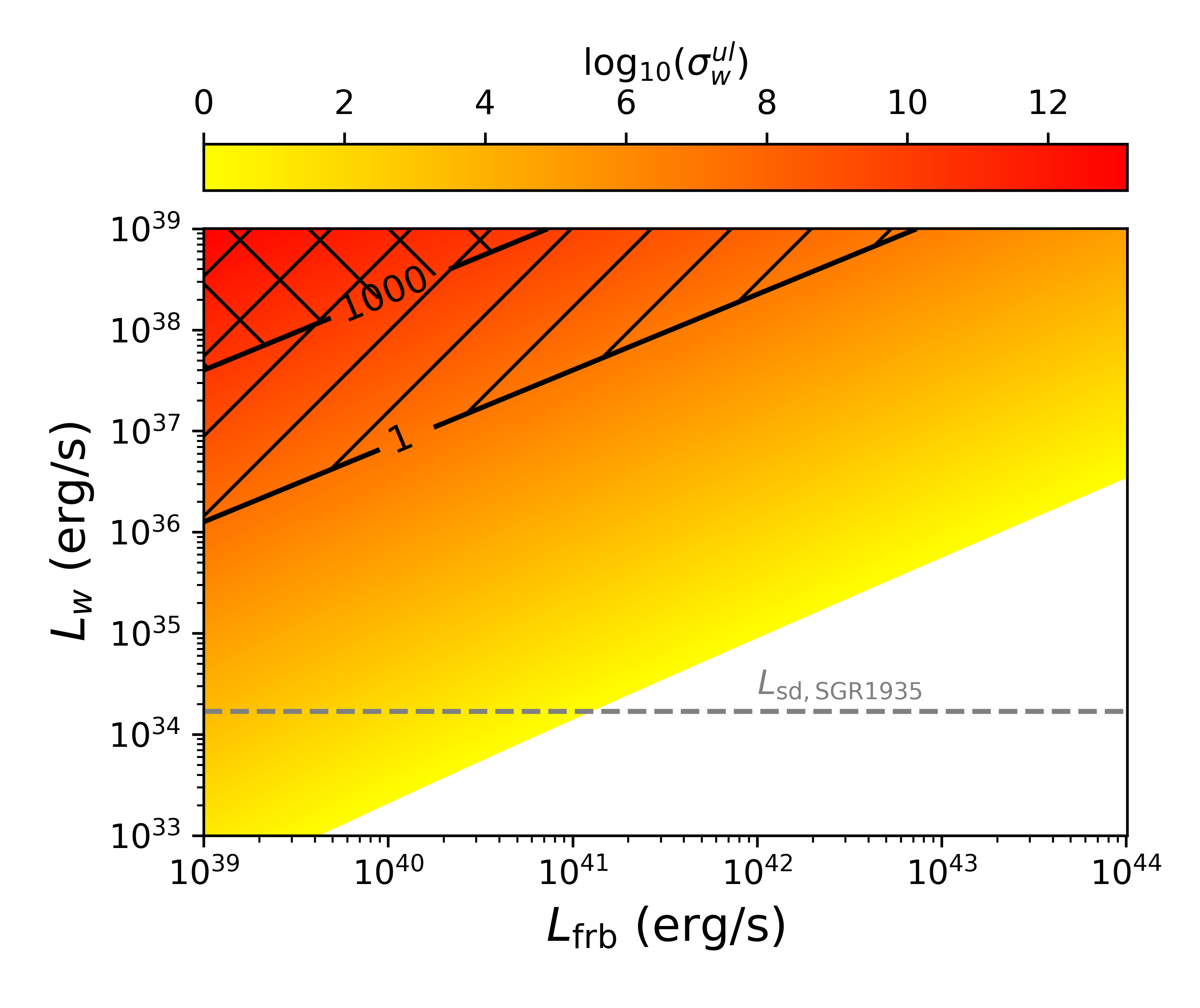}
    \caption{log$_{10}$ of $\sigma_w^{ul}$ --- the upper limit of $\sigma_w$ --- required to generate detectable circular polarization ($|V/U_0| \gtrsim 0.01$), as a function of $L_{\rm frb}$ and $L_{w}$. The white region represents the parameter space that is not allowed, if we assume the wind to be magnetically dominated. 
    We can also use the frequency resolution, $\nu_{\rm res}$, of our detector to additionally rule out the parameter space where $\Pi_V < 0.01$ due to depolarization.
    The black diagonal contours denote that the parameter space toward the top-left is ruled out, in the case where $\sigma_w > \{1, 1000\}$, when $V(\nu)$ is averaged over $\nu_{\rm res} = 1$ MHz. The respective regions are shown as hatched and cross-hatched. Alternatively, the same contours and hatches work with the combination of $\sigma_w > 1$ and $\nu_{\rm res} = \{1 \;{\rm MHz}, 1 \;{\rm kHz} \}$. The contours are an observational limitation derived result, and make no restriction to the physically possible parameter range.
    The horizontal dashed gray line marks the spin-down luminosity of SGR 1935+2154 \citep{Israel2016}, indicative of a typical active magnetar.
    Our analytical model has been used to generate these constraints.
    }
    \label{fig:param_independent_limit}
\end{figure}

We now ask: what region of wind parameter space is compatible with detectable circular polarization? The magnetar wind is expected to be magnetically dominated, with $\sigma_{w0} \gtrsim 1$. If a requirement that $|V/U_0| \gtrsim 0.01$ is imposed, then an upper limit on $\sigma_{w0}$ can be obtained from the analytical model, under the assumption $(1+\sigma_w)^2 / \sigma_w \approx \sigma_w$ when $\sigma_w \gg 1$. This upper limit, labelled $\sigma_w^{ul}$, is shown in Figure \ref{fig:param_independent_limit}. By construction, the factor $(1+\sigma_w)^2/\sigma_w \geq 4$, and it excludes a significant part of the parameter space (shown as the white colored region). No limitation has been imposed on the Lorentz factor ($\gamma \geq 1$).

In addition, we show in Figure \ref{fig:param_independent_limit} that the requirement of measurable circular polarization limits the wind luminosity. Higher wind luminosities and lower FRB luminosities require higher values of $\sigma_w\gamma^2$ to be able to explain measurable $\Pi_V$. For example, if we assume that $\sigma_w \gamma^2 > 1000$, then an FRB with $L_{\rm frb} \sim 10^{39}$ erg/s will have its circular component depolarized below our detection threshold, over $\nu_{\rm res} = 1$ MHz, when $L_w \gtrsim 10^{38}$ erg/s (this point lies in the cross-hatched region in Figure \ref{fig:param_independent_limit}).

\subsection{RM contribution of the wind} \label{sec:discussion_rm}
If the magnetic and rotation axes of the central object are aligned or close to being so, the magnetic field lines in the polar region of the wind are primarily radial ($\mu \gg 1$). In such a case, CP is small. Even though, the primary exchange happens between Stokes $Q$ and $U$, the frequency dependence of PA is not proportional to $\lambda^2$. This is because in the inner part of the wind, where $\alpha \gtrsim 1$, we note that $\eta_1$ does not depend on frequency. The frequency dependence is imprinted only in the outer part of the wind, and it may be confused with $\psi \propto \lambda^2$ only in narrow-band observations with $\Delta \omega / \omega \ll 1$. Only when $r_\alpha < R_0$ and the field lines are almost completely radial can the PA exactly be $\psi \propto \lambda^2$.

In our fiducial case with $\mu \gamma_0 \sim 1$, generalized Faraday Conversion happens most strongly in the inner part of the wind. 
The parameter $\delta$, which represents the tilt of $\vec{\Omega}$ from the $Q-U$ plane, starts out at its maximum value (see Figure \ref{fig:delta_radius}). If $\delta(R_0) \gg 1$, then $\vec{\Omega}$ is almost aligned with $\hat{\bf V}$ and almost pure Faraday rotation should be obtained. But, $\eta_1 \propto L_{w,37}^{3/2}L_{\rm frb,43}^{-1}$ follows the inverse trend of $\delta \propto L_{w, 37}^{-1/2} L_{\rm frb,43}^{1/2}$ when $\alpha \gg 1$ (from Eq \ref{eq:eta1_full}). Consequently, if $\delta(R_0)$ is higher, $\eta_1$ and correspondingly the extent of Faraday rotation are smaller.

On the contrary, when $\eta_1$ is large, so is $\eta_2$, which gives rise to significant CP.
Since, pure $Q-U$ mixing is not happening, $\int \eta_1 \cos \theta_B' dr$ does not yield an RM, and
the information about the exchange between Stokes Q and U is lost within the general evolution of the polarization state. Since, $\delta$ has no dependence on $\omega$ for $\alpha \gg 1$, therefore, $\vec{\Omega}$ for all frequencies move towards the $Q-U$ plane together in the inner wind.
Far away in the wind, the magnetic field becomes almost toroidal. 

Hence, the physical conditions of our model lead to a negligible contribution to the total RM budget, which has been checked with the \textit{RM-Synthesis} method (see Appendix \ref{app:frb20180301a}).
Consequently, the region of large $V$ contribution, that is, the compact object wind, must be different from any region of large RM contribution, with the latter assumed to be associated with the wind nebula/PRS \citep{Yang2020, Yang2022} or a supernova remnant. 

In the Faraday mixing scenario of \citet{Wang2025}, the background B-field can be chosen in an arbitrary manner to allow both Faraday Rotation and Conversion to happen in the same region, which is not possible for our model. In contrast, \citet{Gruzinov2019} require a Faraday rotation dominated region interspersed with magnetic reversal points, otherwise the relationship $V \propto {\rm RM}^{1/2} \lambda^2$ breaks down, and we get complex polarization evolution.

\subsection{Comparison with \citet{Lyutikov2022} and application to Galactic magnetars/pulsars} \label{sec:lyutikov2022}

\citet{Lyutikov2022} also deals with the generation of large circular polarization and large RM in the electron-positron wind of magnetars, but in the regime where $\omega'$, $\omega_p' \ll \omega_B'$. The cyclotron resonance radius, defined where $\omega' = \omega_B'$, is
\begin{equation}
    r_B = (1.2 \times 10^9 \;{\rm cm}) B_{\rm ns, 12}^{1/3} R_{\rm ns, 6}, \omega_{10}^{-1/3}, 
\end{equation}
where, $B_{\rm ns}$ is the surface magnetic field strength in Gauss, and $R_{\rm ns}$ is the neutron star radius in cm.
But, for typical Galactic sources, $R_0 = (4.7 \times10^9\;{\rm cm}) (P/1\,{\rm s})$, which means that generally, $r_B < R_0$. For SGR 1935+2154, $r_{B, \rm SGR1935} = 7.3 \times 10^9$ cm, using $B_{\rm ns, SGR1935} = 2.2 \times 10^{14}$ G \citep{Israel2016}. Correspondingly, $R_{0, \rm SGR1935} = 1.5 \times 10^{10} \; {\rm cm} > r_{B, \rm SGR1935}$. Therefore, the regime of \citet{Lyutikov2022} is not valid for SGR 1935+2154 bursts reported in \citet{Kirsten2021} and discussed here in Section \ref{sec:data_analysis}. Eq (\ref{eq:chiinf_nonrel}) holds in the wind of SGR 1935+2154 and other typical neutron stars and magnetars; $r_B > R_0$ only for rapidly spinning young neutron stars, like the Crab pulsar. 

Moreover, the Cotton-Mouton effect for $\omega' \gg \omega_B'$ is an inherent part of our formalism\footnote{For $\omega' \gg \omega_B'$, the Cotton-Mouton effect arises from the quadratic term of $B_0$ in a uniform plasma, whereas our formalism is based on a spatially-varying medium.}.
A focused work based on Eq (\ref{eq:chiinf_nonrel}) using well-resolved polarization data from Galactic pulsars and magnetars is required in the small amplitude EM wave regime. Such a work can bridge the gap between FRBs, and Galactic magnetars and pulsars, in terms of near-source propagation-induced polarization effects.

\subsection{Time variability of the wind}
The model proposed in this paper can only explain the spectral, but not the temporal properties of Stokes parameters. The latter are explained by way of changing wind conditions, either because of the magnetar's normal variability or due to energetic activity. This variance can be on a range of timescales, with short X-ray bursts dominating small timescales between roughly 10ms to 100ms (e.g., \citealt{Xiao2023}). 
The observable variability time of wind properties is $\Delta t_{\rm var} \gtrsim r / 2c \gamma^2$ ($\sim 16$ ms at $r=10^{11}$ cm and $\gamma=10$), which is short enough to introduce substantial variation between bursts. Assessing whether the variability is intrinsic to the emission mechanism or not is beyond the scope of this work.

\subsection{The effects of magnetospheric propagation} \label{sec:magnetospheric_propagation}
This work does not treat the case of FRB propagation in the magnetosphere because of several unknowns.
First, the magnetospheric B-field may be quite perturbed due to magnetar activity. Second, the plasma response to the high $\alpha$ EM pulse is complicated by the strong background magnetic field, reducing the effective value of $\alpha$ \citep{Lyutikov2024}. Third, the ponderomotive force exerted by the pulse head can displace the plasma, both laterally as well as longitudinally. The plasma particles carried longitudinally by the head of the wave may penetrate only a small fraction of the pulse (the so-called snowplow effect; see e.g., \citealt{Decker1996, Pukhov2002, Lyutikov2024}). This effect is not captured in our simple treatment of the relativistic mass increase effect. 

The snowplow effect decreases as $\alpha$ decreases, and after a certain point
the particles are able to pass through the wave \citep{Sazegari2006}. This effect may also depend on the ion loading of the wind. As long as snowplow is active, the polarization properties of the FRB's main pulse may not be affected. Particle-in-cell simulations can inform where snowplow effects become small (see e.g., \citealt{Sridhar2026}). This can provide constraints for cases where magnetospheric contribution may be required, such as for Burst 926 in \citet{Xu2022}. In general, any magnetospheric propagation model should also satisfy the condition of not being able to induce circular polarization in a majority of bursts.

\subsection{Intrinsic circular polarization}
Our model is not able to distinguish between intrinsic and extrinsic CP contributions. It is possible to fit a Stokes-$V$ spectra with our model even when the underlying mechanism of generating circular polarization could be entirely intrinsic. In many of the bursts we have discussed in this work, Faraday conversion in the wind is able to reasonably explain the observed circular polarization, without requiring an initial non-zero Stokes-$V$. However, not all bursts can be fit this way, and some, like the burst identified as P05-C1 from FRB 20201124A \citep{Kumar2022, Kumar2023}, require an initial non-zero Stokes-$V$. 

Furthermore, for repeaters which consistently exhibit zero CP in hundreds of bursts, and have high CP in only one or two bursts, it seems more reasonable to attribute this behavior to intrinsic mechanisms. Faraday conversion can explain repeaters which exhibit a variety of CP properties, like FRB 20201124A, FRB 20220912A, and FRB 20240114A. 

\subsection{Reasons for not using FRBs 20121102A and 20190520B} \label{sec:plasma_scattering_discussion}
We do not use some active repeaters with long-term monitoring campaign data, like FRB 20121102A and FRB 20190520B. 
Both sources are seen to be fully linearly polarized at high frequencies. 
However, both exhibit close to complete depolarization at typical observing frequencies $\sim 1-2$ GHz attributable to RM scattering caused by multipath propagation \citep{Feng2022a}, with some bursts exhibiting CP in the L-band \citep{Feng2022b}. Due to multipath propagation, burst depolarization and induction of circular polarization can happen for screen rotation measures RM$_{\rm screen} > 10 \nu_9^2$ \rmunit \citep{BKN2022}. Additionally, these authors argue that CP should be more prominent at lower frequencies. Both these conditions are satisfied by the FRBs in question, which points to CP at low frequencies being generated by mechanisms besides Faraday Conversion.

\section{Summary} \label{sec:summary}

In this paper, we aim to determine if circular polarization in FRBs can be produced extrinsic to the source in a physically realizable environment, such as the relativistic wind from magnetized compact stellar objects. We find that such an environment can contribute high levels of circular polarization, but we are not able to rule out the possibility of contributions from alternate sources, such as the intrinsic emission mechanism, the magnetosphere of these compact objects, or the wind nebula at farther distances.

\subsection{Summary of model expectations}

We find that the parameter which controls the extent of circular polarization and its spectral variation, for a given FRB luminosity, is a combination of wind properties $\xi_E \equiv L_{w,37}^2\sigma_{w0}m / m_E(1+\sigma_{w0})^2 \gamma_0^2$; where \(L_w\), \(m_E\) \(\sigma_{w0}\), and \(\gamma_0\) are the wind luminosity, the effective mass of the wind particle, the wind magnetization parameter at the base of the wind, and the Lorentz factor also at the base, respectively.

The level of spectral variation in the Stokes parameters, or a lack thereof, helps accurately determine wind conditions when there is measurable CP.
If there is negligible spectral variation, high $V$ can be achieved at comparatively low values of $\xi_E \approx L_{w,37}^2 m/ m_E\sigma_{w0} \gamma_0^2$. The same is seen for the 90.9\% circularly polarized burst of FRB 20201124A (\citealt{Jiang2024}; also expected for similar bursts, like Burst number 1472 from \citet{Xu2022} with 75\% CP). 

Faraday conversion in magnetar winds naturally explains why most bursts exhibit negligible circular polarization, as the wind parameter \(\xi_E\) is generally small.
For higher luminosity FRBs, the wind parameter needs to be higher to explain the same level of circular polarization, provided there is no intrinsic contribution.
When the wind parameter is extremely high, it can cause detector-level depolarization of the circular component, as the scale of spectral variation becomes smaller than the frequency resolution or the bandwidth over which average polarization is calculated.
Linear components are not depolarized, whereas Stokes $V$ is (in most scenarios). In the case where the frequency resolution $\nu_{\rm res} \ll \nu$, we show that the fractional circular polarization, $\Pi_V \propto \nu / \xi_E \nu_{\rm res}$ at high $\xi_E$ (or high wind luminosities). Consequently, $100\%$ polarized bursts can become partially depolarized. 

In summary, we have shown that there is no CP at low $\xi_E$; it increases in magnitude as $\xi_E$ increases but with negligible frequency dependence. With further increase in $\xi_E$, CP starts oscillating with frequency, and then depolarizes as the oscillations become too rapid for the detector's resolution (see Figure \ref{fig:piV_vs_K_nu_analytical}). 

\subsection{Summary of unique results from data analysis}
Through the analysis of FRB 20180301A, this work raises questions about inferring propagation conditions from empirical fitting models, such as the Generalized Faraday Rotation (GFR) model \citep{Uttarkar2024}.
We establish that the discovery burst's RM$_{\rm obs} = -3163$ \rmunit \citep{Price2019} is within the $68\%$ confidence interval of the RMs required in addition to our model, rather than the GFR-derived RM of 28 \rmunit \citep{Uttarkar2024}. This supports a key model consequence that significant circular polarization must arise in a region separate from where a large RM originates, if the burst exhibits both.

Non-zero circular polarization may still be generated in the wind when the radio luminosity is low such that the dimensionless wave amplitude is less than unity at the base of the wind. We use this case to infer the wind properties of SGR 1935+2154 in Section \ref{sec:data_analysis} (and Appendix \ref{sec:frb20200428}), obtaining $\sigma_{w0}\gamma_0^2 \geq 3 \times 10^4$ and the proton loading to be $\lesssim 1\%$.
The application of this regime can be extended to low luminosity bursts, from sources like FRB 20200120E (the M81 globular cluster FRB) or other Galactic radio sources.

This model cannot explain the Stokes-$V$ spectral variations of burst 926 of FRB 20201124A from \citet{Xu2022}. We propose a heightened ion fraction as well as significant reorganization of the wind magnetic field due to outburst activity to explain observed properties, provided there is no intrinsic or magnetospheric explanation. We also highlight that the field reversal model of \citet{Gruzinov2019} combined with an initial circular polarization can explain this burst.

We emphasize the importance of collecting spectro-polarimetric data of FRBs, and their potential in providing detailed information about near-source environments. Statistically significant samples of individual repeaters and of non-repeating FRBs can be helpful in untangling some aspects of the enigma that are FRBs.

\bigskip\bigskip
\noindent {\bf ACKNOWLEDGMENTS}

\medskip

A significant part of this work was completed while PK was visiting the University of British Columbia. He is grateful to Jeremy Heyl for his hospitality and for the opportunity to work on such a beautiful and inspiring campus. OG is thankful to Jinchen Jiang, Heng Xu, Pavan Uttarkar, KJ Lee, Pravir Kumar, Franz Kirsten, Kenzie Nimmo, and Apurba Bera for providing easy-to-use FRB data and calibration code. OG also thanks Pavan Uttarkar, Zorawar Wadiasingh, Ioannis Contopoulos, and Marcus Lower for helpful discussions, and the organizers of the FRB 2025 conference where several of these discussions were facilitated. OG appreciates Owen Caleb Chase for providing feedback on presentation and writing.

The authors acknowledge the Texas Advanced Computing Center (TACC) at The University of Texas at Austin for providing HPC resources through project AST22018. The authors also acknowledge the use of Large Language Models such as Google Gemini and OpenAI's ChatGPT, which have been used to conduct searches of circular polarization data and literature related to FRBs and pulsars, along with use in improving text readability. The authors also acknowledge that UT Austin, where this research has been conducted, is on the Indigenous land of the Tonkawa people, and historically, the Comanche and Apache moved through this area. This work was funded in part by the National Aeronautics and Space Administration (NASA 80NSSC24K0770; PK and PB). OG was funded by the NASA grant as well as the J. Craig Wheeler Endowed Fellowship in Astronomy from UT Austin for part of the project's duration. PB is also supported by a grant (no. 2024788) from the United States-Israel Binational Science Foundation (BSF), Jerusalem, Israel, by a grant (no. 1649/23) from the Israel Science Foundation.

\software{PSRCHIVE \citep{psrchive2012}, RM-Tools \citep{VanEck2026_rmtools}}

\vfill\eject

\appendix

\section{Transforming polarization transfer coefficients to the lab frame} \label{app:poltransfer_transform}
The polarization transfer Eqs (\ref{dP-dz1}) - (\ref{eq:Omega_def}) and Eq (\ref{dP-dz4}) are dependent on coefficients $\eta_1$ and $\eta_2$, defined in Eqs (\ref{eq:eta1_OmegaV}) and (\ref{eta2-deflab}) respectively.
Their definitions are provided using lab frame quantities, but they follow their standard definitions in the wind frame. $\eta_1$ is the Faraday rotation coefficient, or the first-order coefficient in $\omega_B/\omega$ denoting the difference in the refractive indices of right- and left-circularly polarized light propagating along the background magnetic field. When integrated along the line-of-sight, it gives the polarization angle $\psi$. $\eta_2$ is the Faraday conversion coefficient, and represents the rate at which the phase difference between perpendicular electric field vectors accumulates. Here, we will derive the lab frame $\eta_1$ and $\eta_2$ based on the review article by P. Kumar \& P. Beniamini (in prep.).

In the wind frame,
\begin{equation}
    \eta_1' = \frac{\omega_p'^2 \omega_B'}{c \omega'^2}, \quad \quad \eta_2' = \frac{\omega_p'^2 \omega_B'^2}{2c\omega'^3},
\end{equation}
such that, $\Omega_Q' = - \eta_2' \sin^2 \theta_B' \cos 2\phi_B$, $\Omega_U' = - \eta_2' \sin^2 \theta_B' \sin 2\phi_B$, and $\Omega_V' = \eta_1' \cos \theta_B'$. Note that we are using $\phi_B' = \phi_B$ because we assume that $\vec{\beta} \parallel \mathbf{\hat{k}_0}$. Since, $n' = n/\gamma$,
\begin{equation}
    \omega_p'^2 = \frac{4\pi n'e^2}{m} = \frac{4\pi n e^2}{\gamma_0 m} = \omega_p^2
    \label{eq:omegap_transform}
\end{equation}
The parallel B-field gets transformed as $B_{0\parallel}' = B_{0\parallel}$. In the wind frame, there is no residual force on the particles: the wind frame electric field is zero. Consequently,
\begin{equation}
    \vec{B}_{0\perp} = \gamma(\vec{B}_{0\perp}' + \frac{1}{c}\vec{\beta}\times \vec{E}') = \gamma \vec{B}_{0\perp}'.
\end{equation}
Then $|B_0|^2 = \gamma^2B_{0\perp}'^2 + B_{0\parallel}'^2 = \gamma^2(B_0'^2 - (\vec{B}_0'\cdot\hat{\boldsymbol{\beta}})^2) + (\vec{B}_0'\cdot\hat{\boldsymbol{\beta}})^2 = \gamma^2B_0'^2 - (\gamma^2-1)(\vec{B}_0'\cdot\hat{\boldsymbol{\beta}})^2$. As long as the parallel B-field component is small, $B_0 \approx \gamma B_0'$. Then,
\begin{equation}
    \omega_B' = \frac{eB_0'}{mc} = \frac{e B_0}{\gamma_0 mc} = \omega_B.
    \label{eq:omegaB_transform}
\end{equation}
We also know that $\omega' = \omega / \mathcal{D}$, where $\mathcal{D} = [1 - \vec{\beta}\cdot \mathbf{\hat{k}_0}]^{-1}$ is the Doppler factor \citep{RybickiLightman1986}. Similarly, the distance covered in a relativistically moving cold plasma is decreased because of the motion of plasma itself, leading to $\delta z' = \delta z / \mathcal{D}$.

In the special case when the wind velocity $\vec{\beta} \parallel \mathbf{\hat{k}_0} \equiv \hat{\bf z}$, it can be shown that the Stokes polarization vector in the wind frame $\vec{P}' = \vec{P} /  \mathcal{D}^2$ \citep{Cocke1972}. For $\hat{\boldsymbol{\beta}}\cdot\mathbf{\hat{k}_0} \rightarrow 1$, it is also easy to show that $\mathcal{D} \approx 2 \gamma$. Since the EM wave interaction is most naturally defined in the wind comoving frame, the infinitesimal polarization increment is $\delta P_i' = \varepsilon_{ijk} P_j' \Omega_k' \delta r'$, using Eq (\ref{dP-dz1}), where the cross-product has been represented in the Einstein notation. The Stokes component transformation affects $\delta P_i'$ and $P_j'$ in the same way, thereby cancelling out, and the same equation becomes $\delta P_i = \varepsilon_{ijk} P_j \Omega_k' \delta r' \equiv \varepsilon_{ijk} P_j \Omega_k \delta r$. So, we start out with
\begin{align}
    \Omega_Q' \delta r' &= - \eta_2' \sin^2 \theta_B' \cos 2\phi_B \delta r' = -\frac{\omega_p^2 \omega_B^2}{2c \omega^3 \mathcal{D}^2} \sin \theta_B'^2 \cos 2\phi_B \delta r \nonumber \\
    &= -\frac{2 \gamma^2 \omega_p^2 \omega_B^2}{c \omega^3} \sin \theta_B'^2 \cos 2\phi_B \delta r \equiv -\eta_2 \sin \theta_B'^2 \cos 2\phi_B \delta r \equiv \Omega_Q \delta r.
\end{align}
In showing this result, we have used Eqs (\ref{eq:omegap_transform}) and (\ref{eq:omegaB_transform}). Similarly,
\begin{equation}
    \Omega_U' \delta r' = - \eta_2' \sin^2 \theta_B' \sin 2\phi_B \delta r' \equiv - \eta_2 \sin^2 \theta_B' \sin 2\phi_B \delta r \equiv \Omega_U \delta r.
\end{equation}
Finally,
\begin{align}
    \Omega_V' \delta r' &= \eta_1' \cos \theta_B' \delta r' = \frac{\omega_p^2 \omega_B}{c \omega^2 \mathcal{D}} \cos \theta_B' \delta r \nonumber \\
    &= \frac{2\gamma\omega_p^2 \omega_B}{c \omega^2} \cos \theta_B' \delta r \equiv \eta_1 \cos \theta_B' \delta r \equiv \Omega_V \delta r
\end{align}

\section{Understanding the implications of a realistic magnetar wind}

\subsection{Explicit form of $\eta_1$} \label{app:eta1}
\begin{align}
    \eta_1 = \frac{2\psi_p\gamma\omega_p^2\omega_B}{\omega^2c\gamma_{\alpha}^2} \cdot \frac{m}{m_E} &= (2.18\times10^{-8}\;{\rm cm^{-1}}) \frac{\psi_p L_{w,37}^{3/2} \sigma_w^{1/2}m}{r_{13}^3 \omega_{10}^2 (1 + \sigma_w)^{3/2}\gamma_\alpha^2 \gamma^2 m_E} \\
    &=\begin{cases}
        (2.18\times10^{-8}\;{\rm cm^{-1}}) \tilde{\xi} r_{13}^{-3} \omega_{10}^{-2} \psi_p m m_E^{-1} \;\; &{\rm for}\; \alpha \lesssim 1, \\
        (2.13\times10^{-9} \;{\rm cm^{-1}}) \tilde{\xi} L_{\rm frb,43}^{-1} r_{13}^{-1} \psi_p m m_E^{-1}\;\;&{\rm for}\; \alpha\gtrsim 1.
    \end{cases}
    \label{eq:eta1_full}
\end{align}
Here,
\begin{equation}
    \tilde{\xi} = \frac{L_{w,37}^{3/2} \sigma_w^{1/2}}{ (1+\sigma_w)^{3/2}\gamma^2}.
\end{equation}

\subsection{Calculating the effective mass for a mixed species plasma} \label{app:effective_mass}

The number of protons per electrons is denoted by $\psi_p$. Using this, we can write $n_p = \psi_p n_e$ and $n_{e+} = (1-\psi_p)n_e$, which are the number densities of protons and positrons respectively. $n_e$ is the electron number density. We can write the total plasma frequency as,
\begin{equation}
    \omega_{p, \rm tot}^2 = \sum_q \frac{4\pi n_q e^2}{\gamma m_q} = \frac{4\pi e^2}{\gamma m} \Big[ (2-\psi_p)n_e + \frac{m}{m_p}\psi_p n_e \Big] \approx (2-\psi_p) \omega_{pe}^2,
\end{equation}
where the summation is over the three species $q$, and $\omega_{pe}$ is the plasma frequency of just electrons. Now, the effective mass is straightforward,
\begin{equation}
    m_E = \frac{mn_e + m_{e+}n_{e+} + m_pn_p}{n_e + n_{e+} + n_p} = \frac{(2-\psi_p)m + m_p\psi_p}{2}.
\end{equation}
With this, we can revise the calculation of the wind magnetization to
\begin{equation}
    \sigma_w = \frac{B_0^2}{4 \pi n_e[(2-\psi_p)m + \psi_p m_p]c^2\gamma}.
\end{equation}
Substituting this in $\omega_B^2/\sigma_w$ leads to the relationship with $\omega_{p, \rm tot}$, given in Eq (\ref{eq:omegap_effective_mass}). The subscript ``tot'' is omitted to maintain semantic continuity.

\subsection{Evolution of $\vec{\Omega}$ on the Poincare sphere} \label{app:omega_evolution_poincare}

As $\vec{\Omega}$ is constrained is move along a longitude of the Poincare sphere due to our assumption of $\phi_B$ being constant, its evolution can be understood using the parameter $\delta$, defined as the tangent of the angle between the $Q-U$ plane and $\vec{\Omega}$,
\begin{align}
    \delta \equiv \frac{\eta_1 \cos \theta_B'}{\eta_2 \sin^2 \theta_B'} = \frac{\psi_p \gamma_{\alpha} \omega \cos \theta_B'(r)}{\gamma_0 \omega_B \sin^2 \theta_B'(r)} \approx
    \begin{cases}
        1000 \psi_p \Big[ \dfrac{L_{\rm frb, 43}}{L_{w, 37}} \cdot \Big(1 + \dfrac{1}{\sigma_w} \Big) \Big]^{1/2} \dfrac{\cos \theta_B'(r)}{\sin^2 \theta_B'(r)} \quad {\rm for}\; \alpha \gtrsim 1 \\
        312.5 \psi_p \omega_{10} r_{13} \Big[\dfrac{1}{L_{w,37}} \Big(1 + \dfrac{1}{\sigma_w} \Big)\Big]^{1/2} \dfrac{\cos \theta_B'(r)}{\sin^2 \theta_B'(r)} \quad {\rm for}\; \alpha \lesssim 1.
    \end{cases}
    \label{eq:delta_cases}
\end{align}

\begin{figure}
    \centering
    \includegraphics[width=0.4\linewidth]{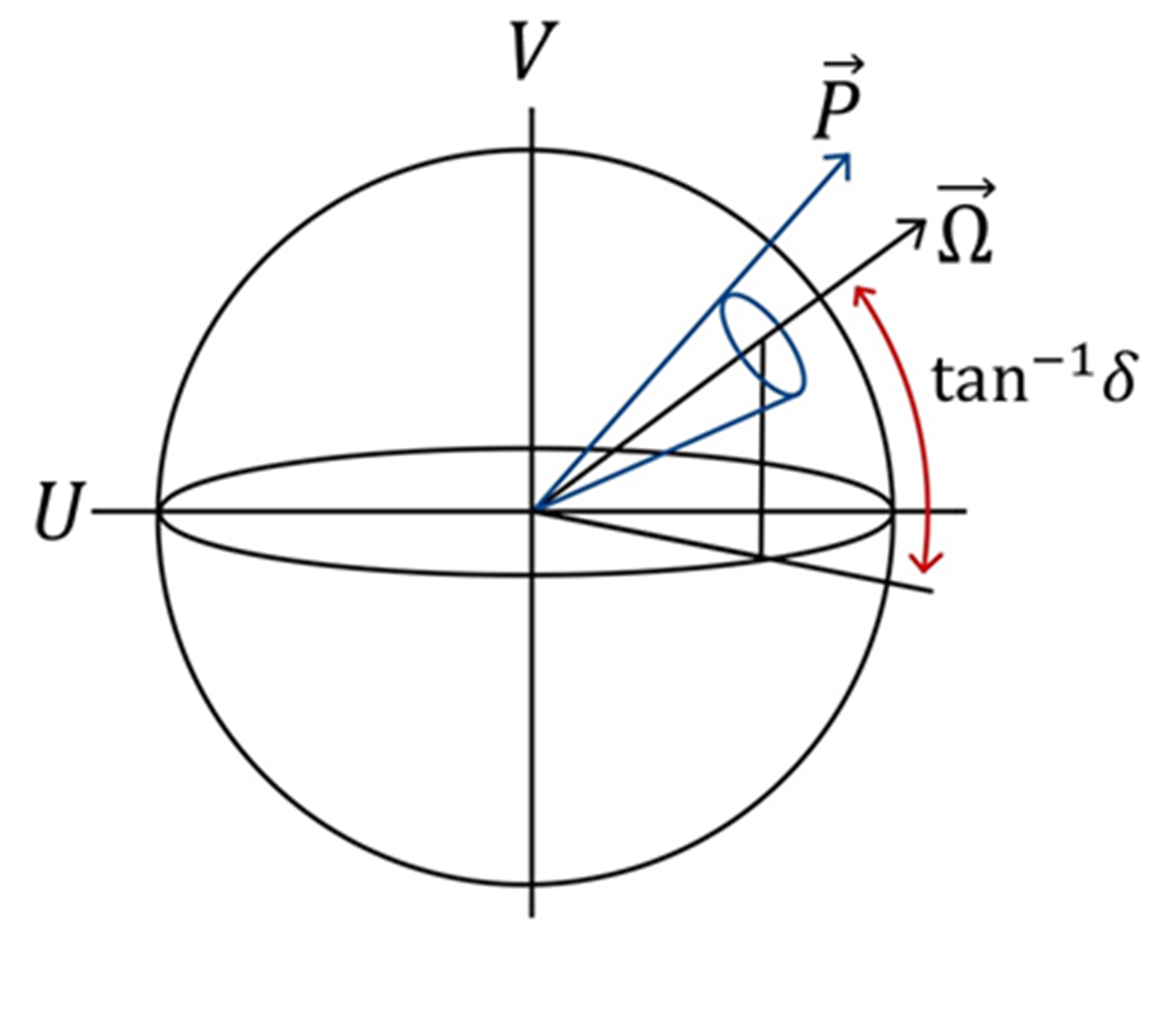}
    \includegraphics[width=0.48\linewidth]{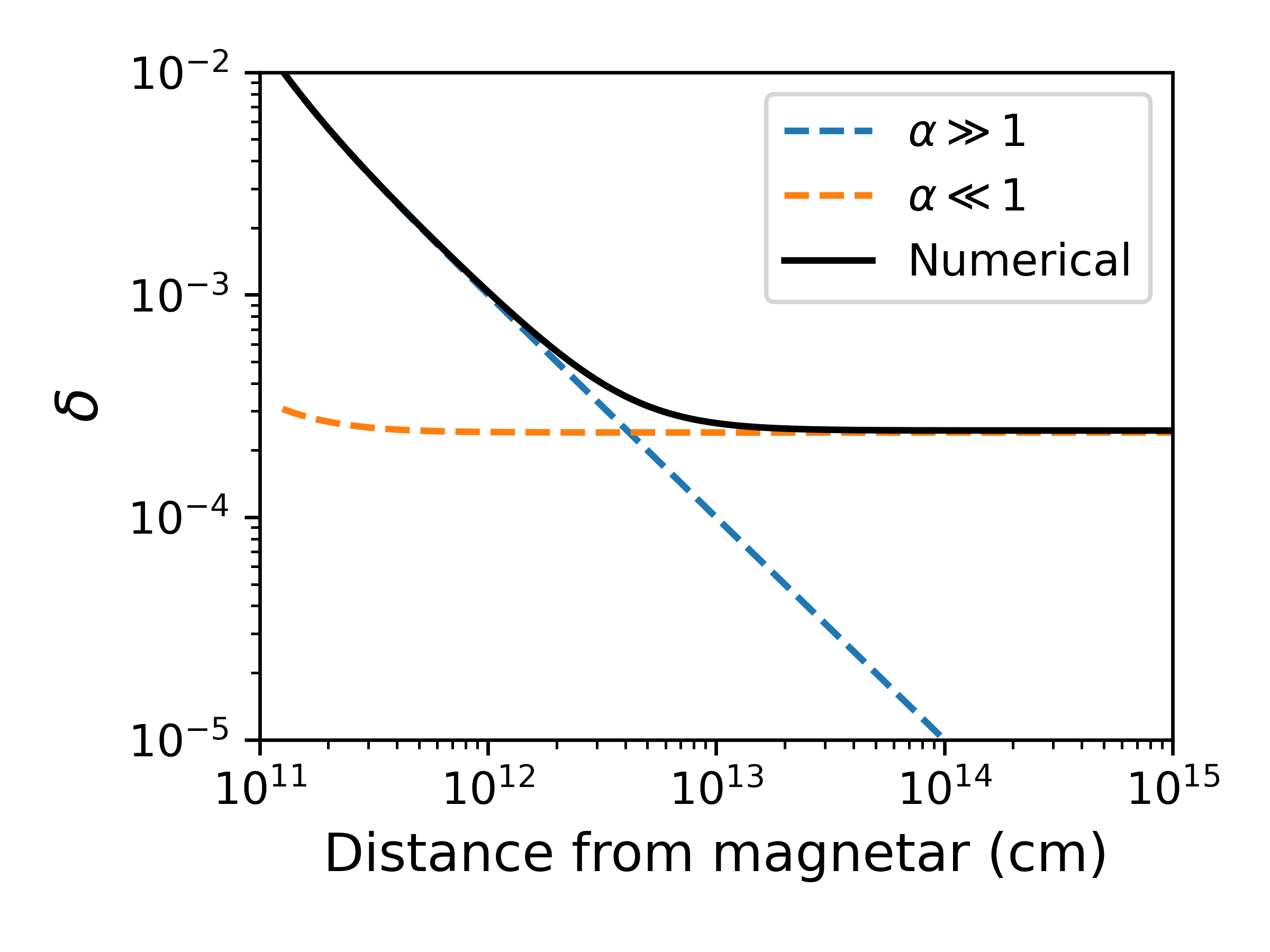}
    \caption{(\textit{left}) A schematic of the Poincare sphere with the $Q$ axis pointing into the page. The polarization vector $\vec{P}$ is depicted as being confined to evolve around $\vec{\Omega}$. The angle between $\vec{\Omega}$ and the $Q-U$ plane is shown to be $\tan^{-1} \delta$. Note that for $\delta \ll 1$, $\tan^{-1} \delta \approx \delta$. (\textit{right}) The evolution of $\delta$ as a function of radius, comparing its numerical and asymptotic analytical regimes. Typically, $\vec{\Omega}$ moves downward in latitude and settles close to the $Q-U$ plane far away in the wind. Radial evolution of $\sigma_w$ and $\gamma$ has been neglected. For this plot, we have chosen $L_{\rm frb} = 10^{41}$ erg/s, $\sigma_{w}=10^4$, $L_{w,37} = 100$, $\gamma = 10$, $\psi_p = 10^{-3}$, $\nu=1200$ MHz, $\phi_B = 15^{\circ}$, and $\vec{P}_0 = (0.7, \sqrt{1-0.7^2}, 0)$.}
    \label{fig:delta_radius}
\end{figure}

It can be simplified further by neglecting the $1/\sigma_w$ term when $\sigma_w \gg 1$, and the corresponding radial evolution is shown in the right panel of Figure \ref{fig:delta_radius}. When $\alpha \gg 1$, the evolution of $\vec{\Omega}$ is significant if FRB luminosity is higher, or if the wind luminosity is lower, or there are more ions in the wind. Alternatively, $\sigma_w$ could be driven to values smaller than 1 to increase $\delta$, which may not be a physically reasonable condition for magnetar wind radii being considered. 

\subsection{Radial evolution of $\sigma_w$ and $\gamma$} \label{app:understand_radial_evolution}

For simplicity, we assume $\sigma_w \gg 1$, such that $1+\sigma_w \approx \sigma_w$. As $\gamma \propto r$ until $R_{\rm fms}$, we define the slope $m_{\gamma} = (\gamma_{\rm fms} - \gamma_0) / (R_{\rm fms} - R_0)$, and consequently,
\begin{equation}
    \gamma(r) =
    \begin{cases}
        m_{\gamma} (r - R_0) + \gamma_0, \quad R_0 < r < R_{\rm fms} \\
        \gamma_{\rm fms}, \quad R_{\rm fms} < r.
    \end{cases}
\end{equation}
Using the conservation of the wind's energy flux, $\sigma_w = \sigma_{w0} \gamma_0 / \gamma$. Then,
\begin{equation}
    \eta_2 = 20 \frac{L_{w,37}^2}{L_{\rm frb, 43}^{3/2}} \cdot \frac{1}{\sigma_w \gamma^2 r} = 20 \frac{L_{w,37}^2}{L_{\rm frb, 43}^{3/2}} \cdot \frac{1}{\sigma_{w0} \gamma_0 \gamma(r) r}.
\end{equation}
This results in,
\begin{equation}
    \zeta_{\infty} = 20 \frac{L_{w,37}^2}{L_{\rm frb, 43}^{3/2}} \cdot \frac{1}{\sigma_{w0} \gamma_0 \gamma_{\rm fms}} \Big[ \frac{\gamma_{\rm fms} -1 }{\gamma_0 - 1} \ln \gamma_0 + \ln \frac{r_{\alpha}}{\gamma_{\rm fms} R_0} + 0.35 \Big].
\end{equation}

It is easily shown that,
\begin{equation}
    \frac{dV}{d\omega_{10}} = - \frac{\cos \zeta_{\infty}}{\omega_{10}} \cdot \frac{20 L_{w,37}^2}{\sigma_{w0} \gamma_0 \gamma_{\rm fms} L_{\rm frb, 43}^{3/2}},
\end{equation}
using which, the ratio of frequency widths over which $|\Delta V| = 1$ between the radially evolving and non-evolving cases, with both having a transverse non-evolving background magnetic field, is $\Delta \omega_{\rm evol} / \Delta \omega_{\rm non-evol} = \gamma_{\rm fms} \gamma_0^{-1} = \sigma_{w0}^{1/3}  \gamma_0^{-2/3}$. This is an important result because the radial evolution of wind properties relaxes the spectral variation by the factor $\gamma_{\rm fms} / \gamma_0$ for the same $\xi$. 

\section{Depolarization of $V$ due to finite frequency resolution of the detector} \label{app:fractional_polarization}
Let us compute the average circular polarization over a small frequency range. We begin with writing Eq (\ref{chi-inf1}) as,
\begin{equation}
    \zeta_{\infty} = \frac{20 \xi}{L_{\rm frb, 43}^{3/2}} \Big[ \ln \Big({ 3.2\times10^{13} L_{\rm frb, 43}^{1/2} \over R_0} \Big) - \ln \Big( \frac{\omega}{10^{10}} \Big) + 0.35 \Big] \equiv a[b - \ln \omega].
    \label{eq:zeta_basic_freq_dependence}
\end{equation}
We find that $d\zeta_\infty / d\omega = - a/\omega$, for the specific example of Eq (\ref{sol1-UV}), the average $V$ over the frequency channel centered at $\omega$ is,
\begin{equation}
    \langle V \rangle_{\rm res} = \frac{U_0}{\omega_{\rm res}} \int^{\omega + \omega_{\rm res}/2}_{\omega - \omega_{\rm res}/2} \sin [\zeta_\infty (\omega')] d\omega',
\end{equation}
where, $\omega_{\rm res}$ is the width of an individual frequency channel. In the limit that $\omega_{\rm res} \ll \omega$, which is valid for FRB observations, we can Taylor expand $\zeta_\infty (\omega') \approx \zeta_\infty (\omega) + (\omega' - \omega) d\zeta_\infty / d\omega = \zeta_\infty (\omega) - a(\omega' - \omega) / \omega$. Additionally, we can substitute $\Delta \omega \equiv \omega' - \omega$. Then,
\begin{equation}
    \langle V \rangle_{\rm res} = \frac{U_0}{\omega_{\rm res}} \int^{\omega_{\rm res}/2}_{-\omega_{\rm res}/2} \sin [\zeta_\infty (\omega) - \frac{a}{\omega} \Delta \omega] d(\Delta \omega).
\end{equation}
Using the identity, $\sin (x-y) = \sin x \cos y - \cos x \sin y$, and noting that the integral of $\sin(a\Delta \omega / \omega)$ is zero, as it is an odd function, we get,
\begin{equation}
    \langle V \rangle_{\rm res} = U_0 \sin \zeta_\infty \frac{\sin K}{K}, \quad {\rm where,} \quad K=\frac{a\omega_{\rm res}}{2\omega} = \frac{\xi_{-1} \nu_{\rm res,6}}{L_{\rm frb,41}^{3/2} \nu_9}.
    \label{eq:avgV_res}
\end{equation}
The full expression is plotted in the left panel of Figure \ref{fig:piV_vs_K_nu_analytical}.

In the above equation, frequency dependence is embedded both in $\sin \zeta_\infty$ and in ${\rm sinc} \, K$. When $K \ll 1$, the frequency dependence goes as $\sin [a(b - \ln \omega)]$. The case-by-case behaviour in different regimes is discussed in Section \ref{sec:fractional_polarization}. It can also be easily shown that $\langle U \rangle_{\rm res} = U_0 \cos \zeta_\infty \,{\rm sinc} \, K$, and consequently $\langle Q \rangle_{\rm res} = \sqrt{1-U_0^2}$ (there is no variation in $Q$ with frequency). In the particular case that is discussed in Section \ref{sec:basic_formalism}, with $Q_0=0,\;U_0=1, \;V_0=0$, both $U$ and $V$ are depolarized.

\section{Analysis of FRB Circular Polarization Data} \label{app:analysis}

In this Section, we compare the observed polarization behaviour of several FRBs, obtaining wind properties for individual bursts. Exact polarization spectrum for FRBs with significant circular polarization can help provide an estimate on $\xi_E$. While such observational data is most useful, most FRBs exhibit small circular fractions, and fitting spectral information is generally overkill. In rare cases, frequency-averaged V/I may yield low values if there is rapid spectral variation. However, most FRBs exhibit negligible frequency-averaged V/I, almost all of which do not exhibit rapid spectral variation (we have only examined a small number of FRB polarization spectra, but we believe the latter claim to be true). For these FRBs, it is not necessary to model each burst's Stokes spectra, and just the pulse-averaged value or an upper limit on the observed circular polarization can give a useful upper limit on $\xi$ in the regime provided $\zeta_{\infty} \leq \pi/2$. The obtained $\xi$ provide novel information about the near-source environment hitherto unknown or unconstrained.

There are certain limitations to the amount of information that can be extracted. The wind model described in this work has too many free parameters, different combinations of which can fit the same spectrum. For example, Eq. (\ref{eq:xi_numerical}) shows a degeneracy between $\psi_p$ and $\xi$. As we have discussed for the numerical model, somewhat different values of $\xi$ can also fit the same spectrum when $L_w$, $\sigma_{w0}$ and $\gamma_0$ are altered. To remove some of this degeneracy, we fix $\psi_p=10^{-3}$ and $\gamma_0=10$, unless otherwise specified. For each spectral fitting, we use two values of $L_w$ in order to understand the range over which $\xi_E$ can be obtained. Furthermore, we will assume that the FRB entering the wind has a flat PA spectral profile and zero circular polarization (the ellipticity angle $\chi_0(\nu) = 0$), unless otherwise specified. This is a reasonable choice to make given that the polarization exiting the magnetosphere is unknown, and highly fine-tuned results need to be avoided.

\subsection{FRB 20201124A} \label{app:frb20201124a}
FRB 20201124A is a highly active repeater with a likely complex magnetized environment. Significant evolution in the RM over several days \citep{Xu2022}, as well as high circular polarization of up to 90\% \citep{Jiang2024} have been discovered. Our model can explain the polarization behaviour of several bursts from this source with the fairly simple wind model.
For this FRB, we use data from \citet{Jiang2024}, who provided detailed spectral and temporal information of bursts with $\Pi_V > 50\%$ and those showing special properties. When performing spectral fitting, care has been taken to select bursts which have a single component burst profile, with the entire emission seen within the observing band. Some bursts with very high V/I do not adhere to these criteria, but have still been selected. For bursts extending outside  the observing band, the FRB luminosity is underestimated, and thus, a higher $\xi$ may be required to explain the obtained polarization, based on Eq. \ref{chi-inf1}.

A chi-squared minimization is performed on the fiducial model's output Stokes parameters converted to the PA $\psi(\nu)$ and ellipticity angle $\chi(\nu)$, for each burst on a uniform grid of $\xi$, $\phi_B$, and $\psi_0$. Both $\phi_B$ and $\psi_0 \in [0, 180)$ with an interval of 2 degrees. The results for all selected bursts is given in Table \ref{tab:frb20201124a_gam0V0}, which are shown for 2 values of $L_{w,37}$ chosen on the basis of the estimated persistent radio source (PRS) luminosity $L_{\rm PRS}$. The PRS associated with FRB 20201124A has an inverted spectrum with $L_{\nu} \propto \nu$, and $L_{\nu}(15 \;{\rm GHz}) = 5.3\times 10^{27}$ erg/s/Hz \citep{Bruni2024}. This yields $L_{\rm PRS} \approx 8\times10^{37}$ erg/s, integrating within the range of observations from $6-22$ GHz. We choose $L_{w,37} = \{1, 100\} \times L_{\rm PRS, 37}$ (see e.g. \citealt{Rahaman2025}). In the chi-squared minimization process, we also obtain all points in the parameter space whose chi-square is less than $1\sigma$ away from the p-value of the minimum chi-square. 

\begin{deluxetable}{cccc} 
\tablecaption{Burst and wind properties for FRB 20201124A bursts.\label{tab:frb20201124a_gam0V0}}
\tablehead{\colhead{Burst} & \colhead{$\langle\Pi_V\rangle$} & \multicolumn{2}{c}{log$_{10}(\xi)$} \\ & & \colhead{$L_{w,37}=8$} & \colhead{$L_{w,37}=800$}}
\startdata
\multicolumn{4}{l}{\citet{Jiang2024}} \\
90 & $0.437$ & $-4.65$ & $-4.63$ \\
123 & $-0.521$ & $-4.86$ & $-4.50$ \\
138 & $-0.626$ & $-4.22$ & $-4.22$ \\
266 & $-0.711$ & $-3.38$ & $-3.31$ \\
298 & $0.621$ & $-3.78$ & $-3.63$ \\
299 & $0.729$ & $-4.47$ & $-4.45$ \\
300 & $0.122$ & $-4.57$ & $-4.56$ \\
398 & $-0.407$ & $-4.18$ & $-4.16$ \\
401 & $0.452$ & $-5.15$ & $-5.26$ \\
497 & $-0.439$ & $-4.27$ & $-4.25$ \\
521 & $-0.909$ & $-5.02$ & $-5.02$ \\
\multicolumn{4}{l}{\citet{Kumar2022, Kumar2023}} \\
P05-C1 & $-0.47$ & $-2.74$ & $-2.05$ \\
P05-C2 & $-0.07$ & $-2.03$ & $-2.00$
\enddata
\tablecomments{$\chi_0=0$ by choice in all except P05-C1.}
\end{deluxetable}

\begin{figure*}
    \centering
    \includegraphics[width=0.4\linewidth]{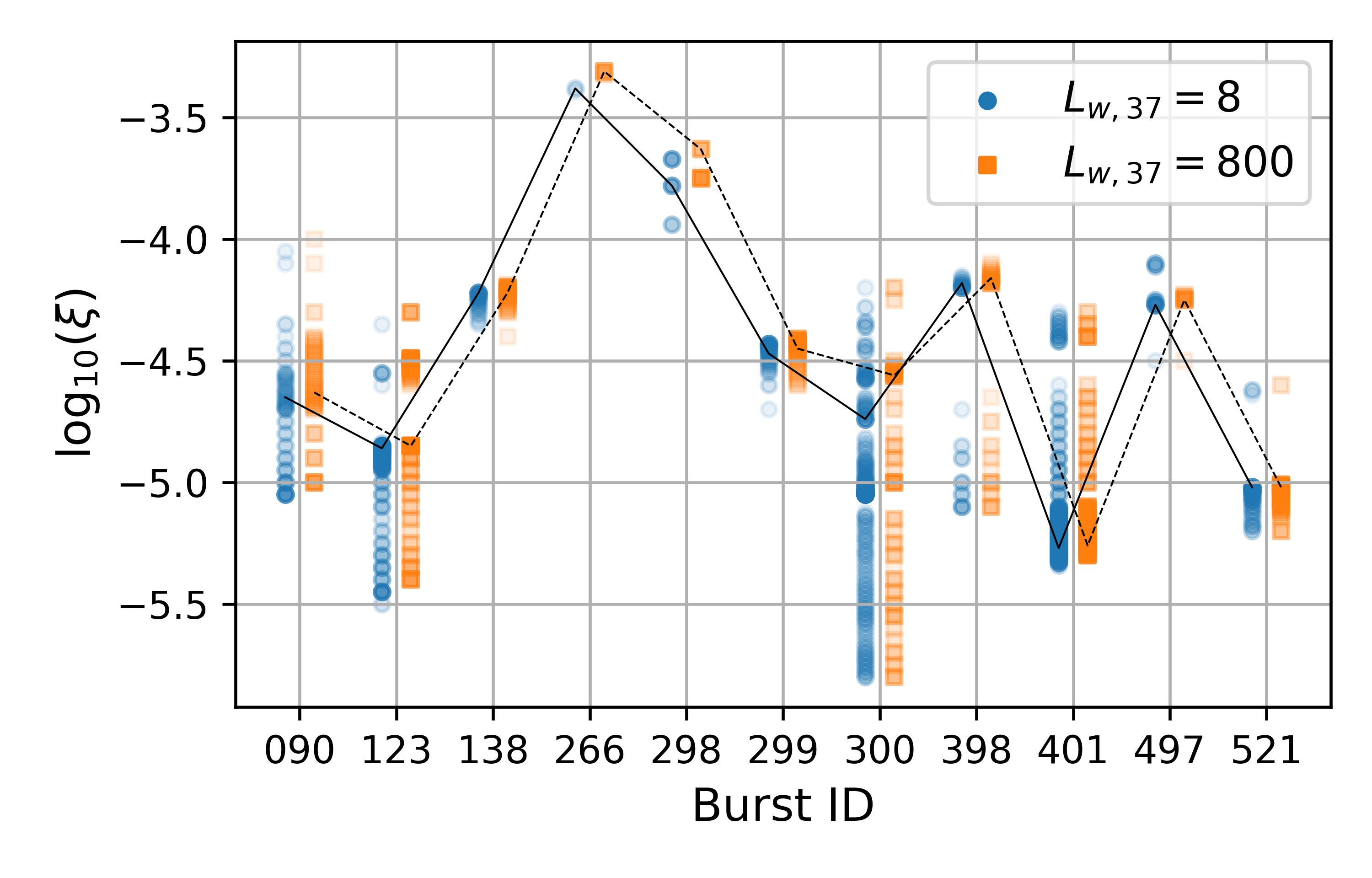}
    \includegraphics[width=0.28\linewidth]{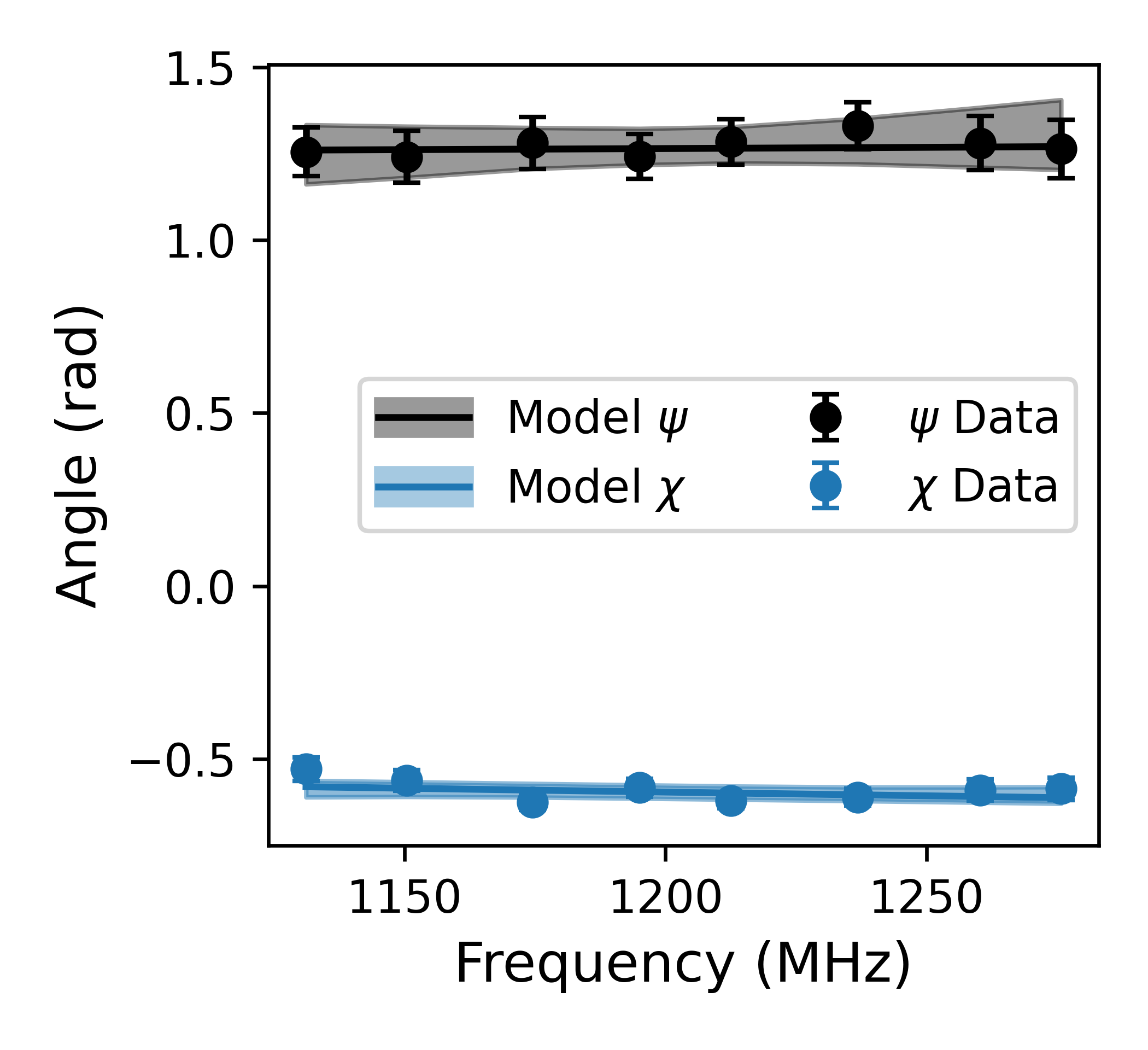}
    \includegraphics[width=0.28\linewidth]{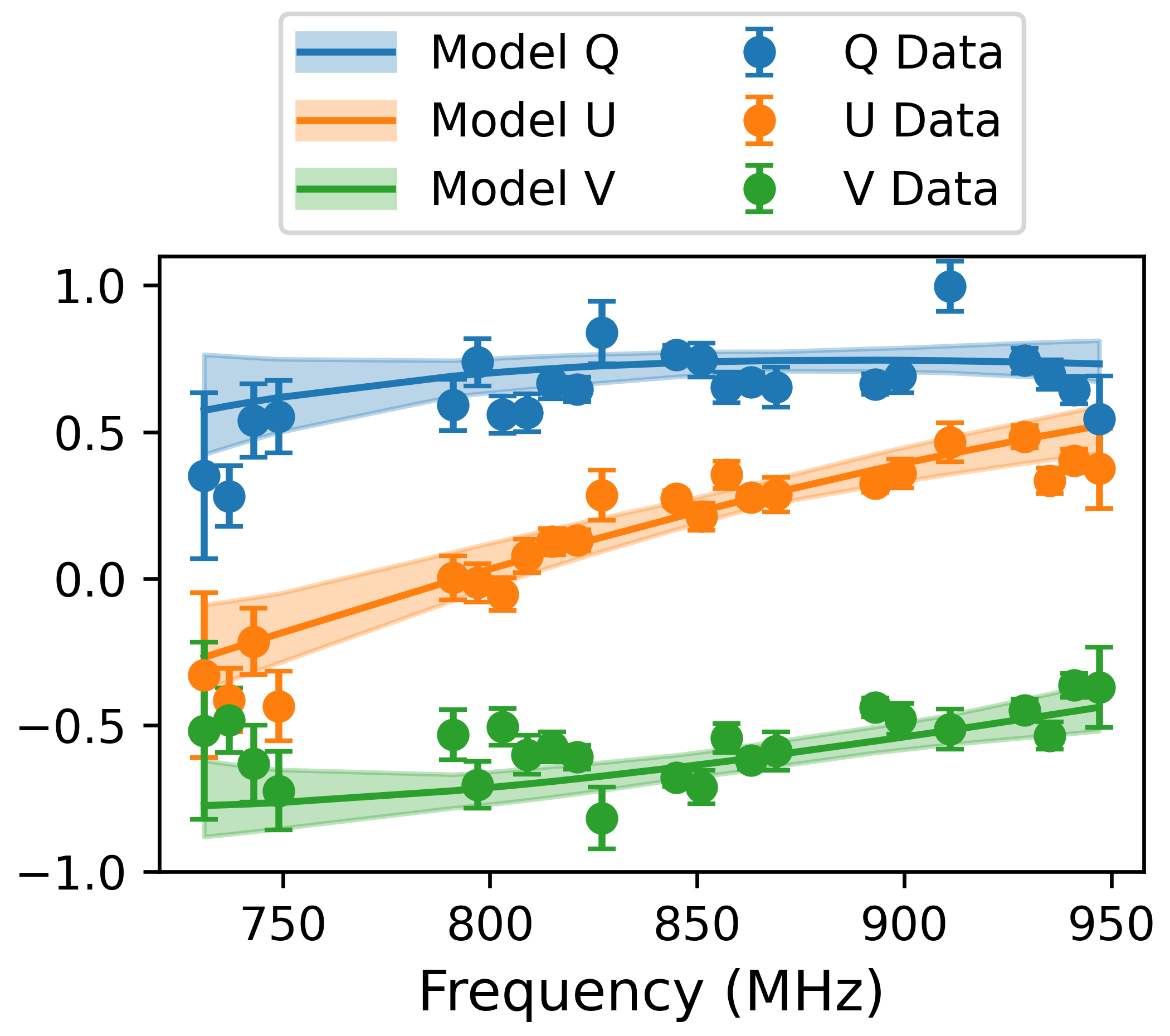}
    \caption{(\textit{left}) The evolution of the best fit wind parameter $\xi$ for selected bursts from FRB 20201124A \citep{Jiang2024}, shown for $L_{w,37} = 8$ (solid black line) and for $L_{w,37}=800$ (dashed black line). Values of $\xi$ which also provide a good fit, within $1\sigma$ of the minimum chi-squared are also shown as colored points (see Section \ref{app:frb20201124a}), with the intensity of color indicating the number of grid points in the parameter space supporting the fit. Other bursts from this observing campaign are not shown, and only bursts $298-300$ are consecutive. (\textit{middle}) Shows the negligible frequency dependence of the polarization angle $\psi$ and the ellipticity angle $\chi$ of the $90.9$\% circularly polarized Burst 521 of FRB 20201124A from \citet{Jiang2024}, and its model fit. This burst highlights that some high-$V$ bursts can have a tiny frequency dependence, which is easily explained by our model but not by previous works. (\textit{right}) The model fit to the Stokes parameters for the first component of the FRB 20201124A burst P05 listed in \citet{Kumar2022, Kumar2023} for $L_{w,37}=8$. This burst requires a non-zero ellipticity angle at the base of the wind to fit the spectrum. The shaded regions show the $1\sigma$ deviation computed in the chi-squared minimization process.}
    \label{fig:frb20201124a_Jiang2024_Kumar2022}
\end{figure*}

We also analyze the highly circularly polarized burst reported by \citet{Kumar2022, Kumar2023}, and coded as P05. This burst has two components separated by a few milliseconds, with the first showing significant CP as well as a strong frequency dependence of polarization. The second component has negligible $V$. We fit the Stokes spectra which are Faraday de-rotated by the measured RM of $-613$ rad/m$^2$ at 896 MHz. The first component requires a non-zero flat $V_0$ spectrum; the corresponding value of log$_{10}(\xi)$ is listed in Table \ref{tab:frb20201124a_gam0V0}, and the fit is shown in Figure \ref{fig:frb20201124a_Jiang2024_Kumar2022}(right panel). The full spectrum in \citet{Kumar2023} has not been fit, as the error in the data becomes significant at higher frequencies. We obtain a residual best fitting RM of $-3$ \rmunit for P05-C1. The second component can be satisfactorily fit with $\chi_0=0$.
Considering that generally $L_w \gtrsim L_{\rm PRS}$, along with assuming $\gamma_0=10$, and taking ${\rm log}_{10}(\xi) \lesssim -2.03$ for burst P05, we obtain $\sigma_{w0} \gtrsim 68$.

\subsection{FRB 20180301A: Rapid frequency-dependent oscillations seen in discovery burst} \label{app:frb20180301a}
FRB 20180301A is also a prolific repeater with ${\rm DM} = 522 \pm 5$ \dmunit and the discovery burst's ${\rm RM} =-3163 \pm 20$ \rmunit \citep{Price2019}. Subsequent RM determinations with FAST in 2019 have yielded an average RM $=542$ \rmunit \citep{luo2020, PravirKumar2023}, which changed sign and showed a roughly linear trend, increasing from RM $\approx-200$ in 2021 to RM $\approx -70$ in 2022 \citep{PravirKumar2023}. The discovery burst was not only an outlier based on its high absolute RM, but also in its spectro-polarimetric properties. The Stokes parameters show rapid variation in frequency, on scales of $\sim 20$ MHz in the source frame, serving as a challenging testbed for theoretical models.

Using a model-independent fitting method for Stokes spectra arising from Generalized Faraday Rotation (GFR; \citealt{lower2021}), \citet{Uttarkar2024} claim that the RM attributable to this burst is only 28 \rmunit, while the rest of the variation in Stokes parameters can be explained by GFR. There is some chance, given that it is a singular burst with high RM and later measurements showing lower RM values, that the true absolute RM of this burst might actually have been lower. To test this within the framework of our Faraday conversion model, we generate several spectro-polarimetric variations and then Faraday rotate them by different amounts. The RM is then recovered using the \textit{RM-Synthesis} technique available in the the \textsc{RM-Tools} Python package \citep{Purcell2020_RMtools}. It is determined that for all except a few cases, \textit{RM-Synthesis} is able to extract a value close to the true RM. The outlier cases clearly show \textit{RM-Synthesis} being unable to converge due to no evidence of a $\lambda^2$ dependence in the PA spectrum.
This leads us to believe that the observationally determined RM is close to its real value for FRB 20180301A. It also shows that our Faraday Conversion model does not yield a distinguishable $\lambda^2$ dependence of the PA, in the most plausible scenarios.

Therefore, we run a simultaneous RM grid search along with the least chi-squared parameter grid search on the non-Faraday corrected polarization spectrum, and impose $\chi_0 = 0$. Since, a PRS for this source has neither been observed nor constrained, we use $L_{w,37}=\{100, 1000\}$; $L_{w,37}=10$ is not used because it yields $\sigma_{w0} < 1$ which is not physically expected. The best fitting values for $L_{w,37}=100$ are: $\xi=-0.025$, $\phi_B=28$ deg, $\psi_0 = 6$ deg, and ${\rm RM} = -3235$ \rmunit. The best fitting values for $L_{w,37}=1000$ are: $\xi=0.45$, $\phi_B=40$ deg, $\psi_0 = 62$ deg, and ${\rm RM} = -3240$ \rmunit. Figure \ref{fig:uttarkar2024_xu2022}(left panel) shows the fit of $\psi$ and $\chi$, i.e., the PA and the ellipticity angle (EA), for the non-Faraday rotated spectrum with $L_{w,37}=100$. 
Furthermore, the RM obtained from our grid search is different from the RM in \citet{Price2019} by $\approx 70$ \rmunit, but our $68\%$ confidence interval includes the observed RM. RM extracting tools find it difficult to recover the correct RM if there is a substantial CP component, which adds uncertainty to the observed RM value.

\subsection{FRB 20220912A}

We also analyze 128 bursts from active repeating FRB 20220912A detected in 1.4 hours of observations between $1.1-1.9$ GHz, with measurable $V/I$ for most bursts upto 58\%, and no preference for low circular polarization \citep{Feng2024}. FRB 20220912A is interesting to study because of its near-zero RM and high $V/I$ \citep{Feng2024}. The host galaxy of this source is identified at a redshift $z=0.0771$ \citep{Ravi2023}.
A PRS non-detection down to $1.8\times 10^{28}$ erg/s/Hz $5\sigma$ upper limit at 1.4 GHz \citep{Hewitt2024} implies a small PRS RM contribution, assuming the PRS luminosity-RM correlation \citep{Yang2020, Yang2022}.
This suggests that the environment imposes a negligible to small RM, and supports the claim in Section \ref{sec:discussion_rm} that CP and RM may result from two separate zones.

\begin{figure*}
    \centering
    \includegraphics[width=\linewidth]{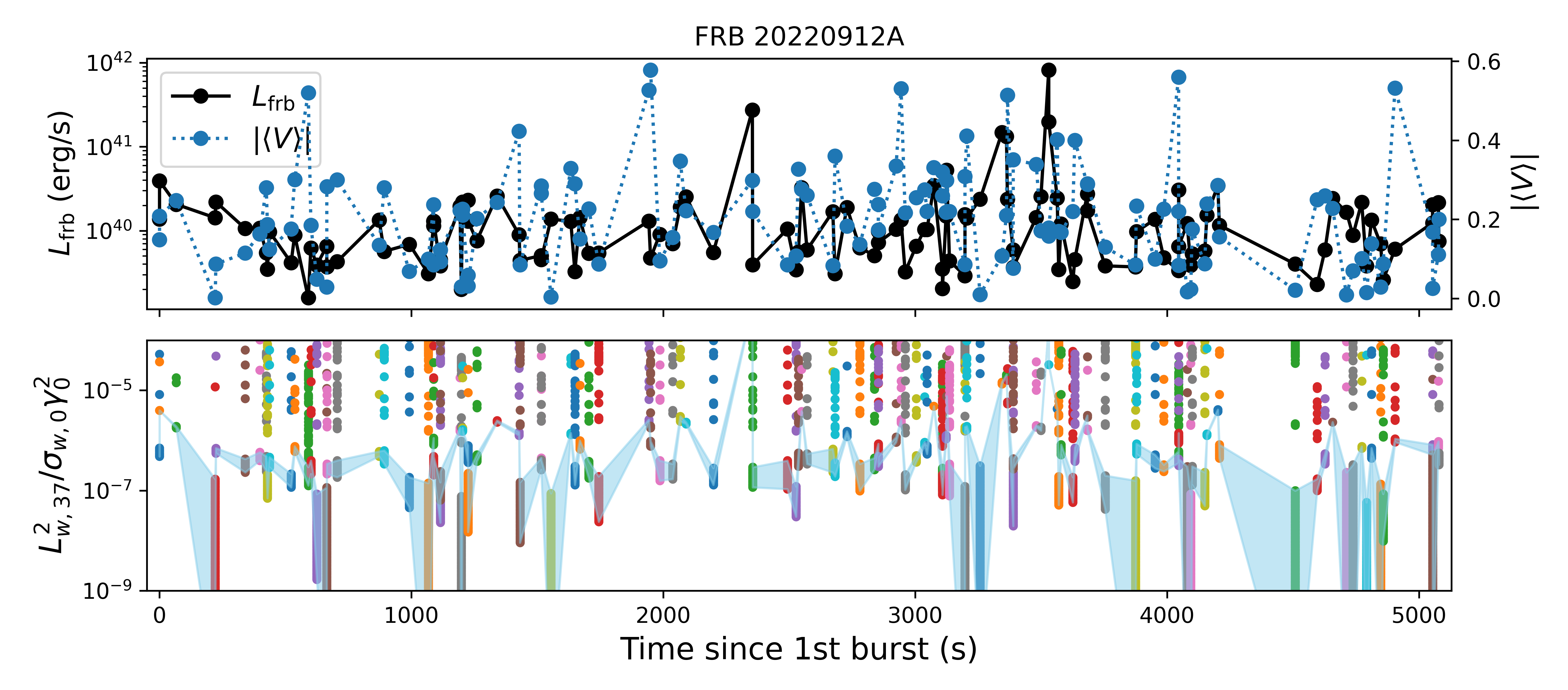}
    \caption{(\textit{top}) Scatter plot of the inferred band-limited isotropic equivalent FRB luminosity and $V/I$ as a function of the arrival time of the 128 burst sample \citet{Feng2024} from FRB 20220912A. (\textit{bottom}) The range of possible values of $\xi \approx L_{w, 37}^2/\sigma_{w0} \gamma_0^2$ giving rise to the observed profile-averaged $V/I$ for the same bursts, assuming $0.1\%$ proton loading per electron in the medium. The colors just distinguish between different bursts. The shaded regions connect the islands of lowest possible $\xi$ values between different bursts, showing a probable time evolution of the wind parameter for this source.}
    \label{fig:frb20220912a_xi_slice_3panel}
\end{figure*}

Spectral width information is not directly available, so we infer it from the isotropic energy and the fluence information computed in \citet{Feng2024}. This is used in conjunction with the peak flux to obtain the isotropic equivalent FRB luminosity. Since we only have $\langle V \rangle$ for all 128 bursts, we cannot fit for the initial polarization. We assume $\vec{P}_0 = (0.7, \sqrt{1-0.7^2}, 0)$, and obtain a range of $\xi$ values which satisfy $\langle V\rangle$, and are shown in the second panel of Figure \ref{fig:frb20220912a_xi_slice_3panel}. This can be related to Figure \ref{fig:Lfrb_iload_vs_xi}, with each burst representing a horizontal slice in the plot at $\psi_p=10^{-3}$. As it cannot be determined which value of $\xi$ is more likely, the lowest $\xi$ satisfying $\langle V \rangle$ are represented by the shaded regions in Figure \ref{fig:frb20220912a_xi_slice_3panel}. The lowest islands of $\xi$ values are chosen because they require a weaker wind parameter to explain the observed circular polarization. It is seen from the lower panel that there is significant variation in the wind parameter. For most bursts, it varies between $\xi \sim 10^{-5}-10^{-7}$, but goes as high as $6\times10^{-4}$. Fitting the Stokes frequency-spectra of bursts would only serve to increase the upper limit for the sample. A similar analysis can be carried out with the significantly larger dataset in \citet{Zhang2023}.

\subsection{FRB 20200428: Low luminosity magnetar bursts showing circular polarization} \label{sec:frb20200428}
Two fairly bright bursts from the magnetar SGR 1935+2154 were recorded with full Stokes information at the Westerbork telescope \citep{Kirsten2021}. While they are not representative of the FRB burst population, and additionally, are more than 4 orders of magnitude dimmer than the MJy-fluence burst from this source \citep{CHIME2020_frb20200428_sgr1935, Bochenek2020}, the relativistic magnetar wind must still be present. The wind affects the propagation of the FRB, even though the EM strength parameter $\alpha \ll 1$ in this region, with $L_{\rm frb, 43} \sim 10^{-10}$. The light cylinder radius for SGR 1935+2154 is $R_{0, \rm SGR1935} \approx 1.55 \times 10^{10}$ cm, inferred from its spin period of 3.245 s \citep{Israel2016}.

The burst separation is $\sim 1.4$s, and data from \citet{Kirsten2021} exhibits frequency-dependent CP. 
The Stokes spectral data is noisy, and additionally, Burst B1 has been suggested to contain 2 burst profiles \citep{Kirsten2021}. We assume that the circular polarization observed arises from Faraday conversion in the wind. Due to the aforementioned issues, instead of fitting the spectra with our model, we just find the parameter space satisfying the frequency averaged $V/P$ across the observing band, which is shown in Figure \ref{fig:sgr1935_contour}. 
 
Results for both bursts obtained using $R_{0, \rm SGR1935} \approx 1.55 \times 10^{10}$ cm in the numerical implementation of Eq (\ref{eq:chiinf_nonrel}), provided in Figure \ref{fig:sgr1935_contour}, show that the bursts occupy a similar region of the parameter space, $\xi \sim (0.15-2)\times 10^{-10}$. This indicates that the wind does not evolve significantly during the window of this observation. Therefore, if the wind is driven by magnetically-dominated outflows, then $\sigma_{w0}\gamma_0^2 \gtrsim 3 \times 10^4$. A further loose constraint on the ion loading of the wind is found to be $\psi_p \lesssim 0.01$ from Figure \ref{fig:sgr1935_contour}.

\bibliography{main}{}
\bibliographystyle{aasjournalv7}

\end{document}